\documentclass[reprint,prl,aps,twocolumn,showpacs,longbibliography,floatfix,10pt,superscriptaddress]{revtex4-2}
\usepackage[dvips]{graphicx}
\usepackage{amsmath}
\usepackage{amssymb}
\usepackage[nice]{nicefrac}
\usepackage{color}
\usepackage{graphicx}
\usepackage{bm}
\usepackage{epstopdf}

\begin{document}


\title{Negative diffusion of excitons in quasi-two-dimensional systems}

\author{A. A. Kurilovich}
\affiliation{Center for Energy Science and Technology, Skolkovo Institute of Science and Technology, Bolshoy Boulevard 30, 121205, Moscow, Russia.}
\affiliation{ Department of Materials Science and Engineering, Technion – Israel Institute of Technology, Haifa 3200002, Israel}
\author{V.~N.~Mantsevich}
\affiliation{Chair of Semiconductors and Cryoelectronics, Physics Department, Lomonosov Moscow
State University, 119991 Moscow, Russia}
\author{A.~V.~Chechkin}
\affiliation{Faculty of Pure and Applied Mathematics, Hugo Steinhaus Center, Wroclaw University of Science and Technology, Wyspianskiego 27, 50-370 Wroclaw, Poland}
\affiliation{Institute for Physics \& Astronomy, University of Potsdam, D-14476 Potsdam-Golm, Germany}
\affiliation{Akhiezer Institute for Theoretical Physics National Science Center "Kharkov Institute of Physics and Technology", 61108, Kharkov, Ukraine}
\author{V.~V.~Palyulin}
\affiliation{Applied AI centre, Skolkovo Institute of Science and Technology, Bolshoy Boulevard 30, Moscow, 121205, Russia.}

\date{\today }
\begin{abstract}
We show how two different mobile-immobile type models explain the observation of negative diffusion of excitons reported in experimental studies in quasi-two-dimensional semiconductor systems. The main reason for the effect is the initial trapping and a delayed release of free excitons in the area close to the original excitation spot. The density of trapped excitons is not registered experimentally. Hence, the signal from the free excitons alone includes the delayed release of not diffusing trapped particles. This is seen as the narrowing of the exciton density profile or decrease of mean-squared displacement which is then interpreted as a negative diffusion. The effect is enhanced with the increase of recombination intensity as well as the rate of the exciton-exciton binary interactions.
\end{abstract}

\maketitle


\section{Introduction}

Excitons are quasi-particles that form when Coulomb-interacting electrons and holes in semiconductors are bound into pair states. These pairs are formed under optical excitation in a wide variety of systems. Among them are inorganic and organic molecular semiconductors \cite{pope1999,colby2010,berghuis2021}, polymers \cite{hadziioannou2000,bolinger2011,kim2007},
semiconductor nanostructures (quantum wells and quantum wires) \cite{ivchenko2005,takagahara2003,scholes2006,grim2014}, colloidal quantum dots and nanoplatelets \cite{nanoplateletschemrev2023, klimov2000,ithurria2011,smirnov2019,smirnov2019_1,Rabouw2016,Shornikova2018,Olutas2015,Brumberg2019,Biadala2014,Meerbach2019}, transition metal dichalcogenides \cite{Mak2010,Splendiani2010,Chernikov2014,Wang2018,Cordovilla2018}, and perovskites \cite{Ishihara1989,Mitzi1994,Becker2018,Belykh2019,Ziegler2020,Magdaleno2021,Seitz2020}. Exciton properties appear to be crucial in a variety of optoelectronic applications including novel lasers \cite{klimov2000}, photodetectors \cite{Konstantatos2012}, light-emitting diods \cite{Baugher2014,Xing2018} and solar cells \cite{Yang2017,Smith2014} since the optoelectronic properties of materials are directly determined by the exciton dynamics. This effects in the growing interest for the transport properties of excitons in semiconductor materials \cite{Yuan2017,Cadiz2018,Kulig2018,Bardeen2014,Dong2015,Kurilovich2020,Kurilovich2021,Glazov2019}. 

The full analysis of exciton transport would require the values of the exciton diffusion length and the diffusion coefficient. Experimentally, it could be done by means of detecting photoluminescence absorption dynamics as a function of pump intensity \cite{Rabouw2016,Shaw2008,Stevens2001,Zamudio2015}, spatially and time-resolved microphotoluminescence  \cite{Kulig2018,Akselrod2014,Han2018,Ginsberg2020} and comparison of photoluminescence signals for semiconductors with present and absent quenching sites \cite{Markov2005,Mikhnenko2008,Kholmicheva2015}. 

Theoretical analysis of exciton transport uses microscopic approaches based on tight-binding models \cite{Kenkre1983,Heijs2005} or kinetic Monte Carlo calculations \cite{Akselrod2014_1,Miyazaki2012}.It can also be based on diffusion equations with a possibility of getting analytical solutions in limiting cases. Diffusion-based models directly demonstrate an important role of traps and static disorder \cite{vlaming2013,Kurilovich2020,Kurilovich2021,Kurilovich2022,Lunt2009,Lee2015,Glazov2019,berghuis2021}. In colloidal nanoplatelets, transition metal dichalcogenides and perovskites the experimental evidence was found that various types of defects could work as exciton traps\cite{Cordovilla2018,Lin2016,Belykh2019,Becker2018,Seitz2020,Kolobkova2021}. The presence of traps splits the population of excitons into two subsystems, the free (mobile) excitons and trapped (immobile) excitons. For the description of their dynamics we earlier have proposed the mobile-immobile model (MIM) \cite{Kurilovich2020,Kurilovich2021,Kurilovich2022}. MIM is a natural extension of the framework of reversible reaction kinetics equations. MIM model captures the physics well in calculations of tracer or contamination spread in geophysical/hydrological applications ~\cite{alekseiralf2022,coats1964,schumer2003}. Also the model was applied in biological systems such as tau-microtubule dynamics~\cite{igaev2014,doerries2022apparent} and as a model of Brownian but non-Gaussian diffusion~\cite{mora2018}. The properties of the model with advection were considered in Ref. \cite{doerries2023emergent}.

While generally MIM assumes that one of the subsets of particles is immobile in real applications a very similar behaviour could obviously be produced by two subsets of particles with a markedly different diffusion constants. This is also often happens with excitons in semiconductors. For instance in transition metal dichalcogenides (TMDs) \cite{Rosati2020,Rosati2021} 
the excitons can change between being in the bright and the dark states. Originally, the laser pulse excites the bright excitons. Some of them recombine and emit photons directly while others transfer into dark excitons. The dark excitons in turn have a relatively high diffusion coefficient right after the transition from the bright state but then thermalise into a "cold" state with much lower diffusion coefficient. The dark excitons and the bright excitons contribute to the same phonon band while the dark excitons also add to another phonon band. This allows to observe the corresponding profiles. The profile for the phonon band $P_1$ then shows narrowing at the intermediate timescales which the authors call as negative diffusion referring to the fact that if one computes the effective diffusion coefficient from the average squared width then the values become negative. Partial negative effective exciton diffusion has been shown also in quantum wells \cite{Zhao2005} and for one-dimensional
quantum molecular chain model due to the competition of phase mixing and diffusion \cite{nakade2021anomalous}.

In tetracene crystals (organic semiconductors) the initial laser excitation results in radiative excitonic singlets which in time split into dark triplets \cite{berghuis2021}. The latter in turn could fuse into delayed singlets as a binary reaction. Only the singlets could be observed in the experiment. The profile of singlet density cross-sections showed a noticeable narrowing at intermediate timescales with widths of cross-sections even smaller than the one produced by the original laser pulse which the authors explain by the non-linearity of the mechanism. Similar singlet-triplet exchange was found to produce the effective negative diffusion in monolayers of WSe$_2$ as well \cite{beret2023nonlinear}. The above mentioned examples show that the negative diffusion usually appears in multi-component systems rather than single-component cases.

One more important phenomenon which strongly affects the dynamics is the exciton-exciton interaction. It occurs when an exciton recombines nonradiatively by transferring its kinetic energy to another exciton.  Being a binary reaction by nature its strength depends on the density of excitons. Fast exciton-exciton annihilation rates typically occur in nanostructures with large exciton binding energies because of the enhanced electron-hole interaction \cite{Philbin2018} hindering their application in optoelectronics. The annihilation was experimentally studied and theoretically analysed in polymer semiconductors \cite{Lewis2006,Stevens2001} as well as modelled extensively \cite{Maniloff1997,Dogariu1998}, however, the exact way the annihilation happens is not yet understood.
\begin{widetext}
\begin{figure}
    \includegraphics[width=1.8\columnwidth]{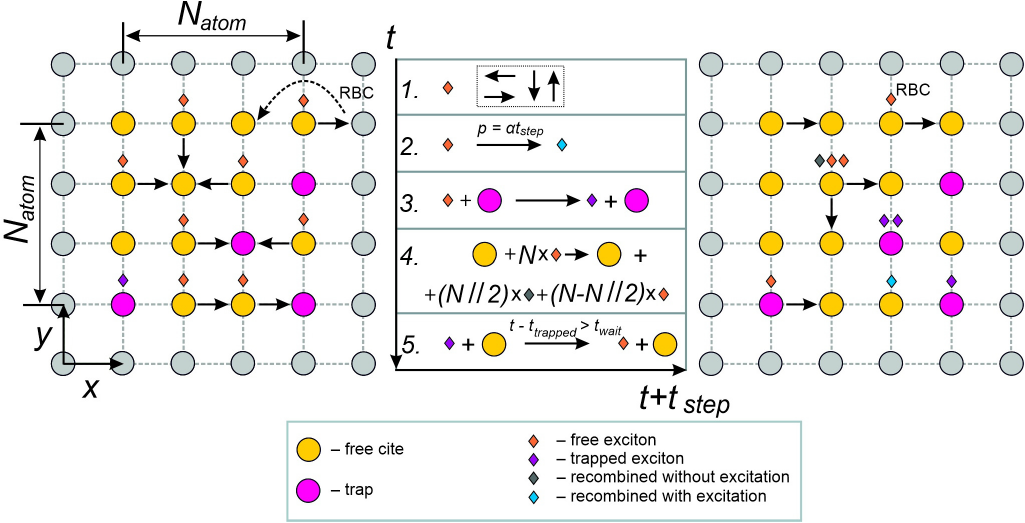}
    \caption{The sketch of the simulation set-up. Excitons perform a simple random walk on a square lattice. They are capable of binary interactions, solitary recombination and trapping. The traps are shown with red vertices. With some probability they are generated at vertices in the beginning of the simulation. The boundary conditions are reflective (RBC).}
    \label{Fig:sketch}
\end{figure}
\end{widetext}
Exciton-exciton interaction causes pronounced non-linear optical effects in quantum wells and colloidal nanoplatelets \cite{grim2014,Baghani2015,Smirnov2021}. In TMDs it was revealed that exciton-exciton interaction strength strongly depends on the surface topography of the substrate \cite{Kajino2021}. The effective exciton-exciton annihilation constant increases with surface roughness, which reflects the fact that locally exciton density focusses due to the corresponding minima of a potential landscape \cite{Hoshi2017}. Investigation of exciton-exciton interaction in the two-dimensional hybrid perovskites demonstrated exciton annihilation rates being more than one order of magnitude lower than those found in the TMDs \cite{Deng2020}, which makes them an exciting class of optoelectronic materials. Thus, exciton-exciton interaction is an important loss mechanism that limits the density of excitons and, consequently, determines the efficiency of solar cells, lasers and LEDs.  

In this article for the analysis of exciton diffusion in a wide class of materials (colloidal nanoplatelets, perovskites, TMDs) taking into account both the presence of traps and the exciton-exciton interaction we propose a simple theoretical  analytically tractable model of a MIM type and a lattice-based simulation model. Both of these models produce a clear understanding for the observable negative diffusion effect and show how the linear and binary recombination enhance it.

The paper is organised as follows. In section II we describe the simulation model and the physical meaning of its parameters. Section III is devoted to results of the simulation model. In section IV we introduce the analytical model and present the results.

\section{Description of models and parameters}
\subsection{Analytical theory}

Based on our previous model \cite{Kurilovich2020,Kurilovich2022} we extend the description of the exciton density evolution for the case with the binary interactions. It is assumed that excitons diffuse in a continuous plane and split into subpopulations: free (mobile) excitons with the density $n_f(\mathbf{r}, t)$ and trapped (immobile) excitons with the density $n_t(\mathbf{r},t)$. The density of all excitons is then obtained by summing the above densities, $n(\mathbf{r}, t)=n_f(\mathbf{r}, t)+n_t(\mathbf{r},t)$. This description corresponds to the framework of the mobile-immobile model (MIM). In our previous papers we showed the usefulness of MIM model for the description of exciton diffusion and PL properties in dichalcogenides, perovskites and semiconductor nanoplatelets \cite{Kurilovich2020,Kurilovich2022}.  We assume that the free excitons diffuse with a coefficient $D$ and could be caught by a trap with the rate $\lambda_f$ while the trapping (escape) times are described by delayed return. This return is modelled by a term with the kernel $\gamma(t-\tau)$, where $t$ is a release time moment as compared to the time of the trapping event $\tau$. The kernel has an explicit
meaning of trapping probability density as a function of time, i.e. it fulfils a normalisation property, $\int_0^\infty \gamma(t)dt=1$. In the case of exponential trapping times the probability density function of trapping simplifies, $\gamma(t)=\lambda_t\exp(-\lambda_t t)$. The distribution of the trapping times with a power-law tail ($\gamma(t\gg\tau_*)\sim\tau_*^\mu/t^{\mu+1},$) could be modelled with Mittag-Leffler distribution (for more details about the distribution and the scaling parameter $\tau_*$ see Appendix \ref{AppendixMittagLeffler}). The linear recombination occurs with the rate $\alpha$ while the binary recombination (annihilation) has a rate $\beta$. Altogether the equations read
\begin{eqnarray}\label{nf}
&&\frac{d n_f (\mathbf{r}, t)}{dt} =-\alpha  n_f(\mathbf{r},t) - \lambda_f n_f(\mathbf{r},t)  \\\nonumber &&+\lambda_f\int_0^t n_f(\mathbf{r}, \tau)\gamma(t-\tau)d\tau - \beta  n_f^2(\mathbf{r},t)+ D\Delta n_f(\mathbf{r},t),\\\label{nt}
&&\frac{d n_t(\mathbf{r},t)}{dt} =\lambda_f  n_f(\mathbf{r},t) -  \lambda_f\int_0^t n_f(\mathbf{r}, \tau)\gamma(t-\tau)d\tau,
\end{eqnarray}
with the initial conditions $n_f (\mathbf{r}, t=0)=N_0\delta(\mathbf{r}),n_t (\mathbf{r}, t=0)=0$. The numbers of free and trapped excitons $N_{f,t}(t) $ can then be obtained by integration of densities,
\begin{eqnarray}
N_{f,t}(t) = \int n_{f,t}(\mathbf{r}, t)d\mathbf{r}.
\end{eqnarray}

The equations (\ref{nf}) -- (\ref{nt}) cannot be solved analytically with a quadratic term in Eq. (\ref{nf}). However, our first-time analytical analysis of MIM model for a particular case with $\beta=0,\alpha\neq0$ provides a substantial insight into the diffusion properties and explains qualitatively the simulation results and some experimental data well.

\subsection{Simulation model}
Our simulation model is lattice-based and generalises our model from Refs. \cite{Kurilovich2020,Kurilovich2021,Kurilovich2022}. Every vertex can be occupied by an exciton. In the absence of interactions the excitons perform a simple random walk with equal jumping probabilities to the nearest neighbours. Some of the vertices could correspond to a low energy state where a particle could be trapped and reside for a while until it escapes. Generally the excitons could recombine linearly or through the interaction with another exciton. In the former case as well as in cases without interactions one could use single particle simulations \cite{Kurilovich2020,Kurilovich2022}. However, in order to simulate exciton-exciton interactions one has to include multiple particles into consideration. In the present work we consider only the binary exciton reactions. It is  assumed that when two excitons collide one of them can recombine while the other one will survive, i.e. it corresponds to the chemical reaction $A+A\rightarrow A$.


The schematic representation of the model with exciton-exciton interactions is shown in Fig. \ref{Fig:sketch}. 
The trap distribution on the square 2D lattice is generated to produce a certain fraction of traps uniformly distributed across the lattice. $N_{0}$ initial positions of the excitons in ensemble are sampled from the discretised 2D Gaussian distribution mimicking the local excitation by a laser pulse. Then the random walk trajectories are produced and analysed. 

Each time step with the duration $t_{\mathrm{step}}$ includes the following stages (Fig. \ref{Fig:sketch}): 

1. $N_f(t)$ free excitons perform a step on the lattice. 

2. Free excitons recombine radiatively with the rate $\alpha$ (ordinary recombination), i.e. at every step the probability of recombination for a free exciton is $p=t_\mathrm{step}\alpha$. The number of these events is then counted and after binning and normalisation can be used to produce a photoluminescence curve.

3. Free excitons arriving to the vertex with a trap are caught for the period of time sampled from a waiting time distribution, in our case, either exponential or Mittag-Leffler distributions. At this stage no trap saturation is assumed thereby simplifying the analysis by removing an additional coupling within the ensemble through the trapped state.

4. Free excitons which happened to be at the same vertex undergo pairwise interactions $A + A \rightarrow A$ with a rate $\beta$.

5. The state of $N_t(t)$ trapped excitons is checked on whether it has to escape the trap or not according to the previously sampled escape times, which are characterised by the rate $\lambda_t$ for the exponential waiting times and by the timescale $\tau_*$ as well as characteristic exponent $\mu$ for the Mittag-Leffler distribution (see Appendix \ref{AppendixMittagLeffler}).

For simulation results, we use the square lattice with reflecting boundary conditions. Some of the semiconductor materials such as TMDs have a hexagonal symmetry. However, as is was shown in our previous work \cite{Kurilovich2022} the results self-average well and no difference between simulations for the square and the triangle lattice were found.

Let us add more details about the parameters used in simulations. The simulations are performed by means of a typical approach used for the excitons dynamics analysis based on the differential equations solution on the lattice \cite{berghuis2021,beret2023nonlinear,Rosati2020}. The $2048 \times 2048$ lattice is used. The probability for the vertex to be a trap is $0.05$. The time discretisation step is $t_\mathrm{step}=10^{-3}\,\mathrm{ns}$. For exponential waiting times $\lambda_t=0.1 \,\mathrm{ns}^{-1}$ is used for the escape rate. For waiting times defined by Mittag-Leffler function the characteristic exponent $\mu=0.65$ and the timescale $\tau_*=5.0 \,\mathrm{ns}$ are used. The $N_0 = 2430$ initial positions are sampled from discretised 2D Gaussian with mean $(1023, 1023)$ and diagonal covariance matrix with $(64, 64)$ at the main diagonal. For averaging $10^3$ different simulations with different random trap distributions and initial exciton distributions are considered except for simulations with annihilation (binary interactions). In the latter case, larger statistics is gathered by performing $3 \times 10^3$ simulations to reduce the noise arising from the rapid decay of non-recombined excitons due to binary interactions. 

We would also like to point out that we do not aim to directly numerically reproduce experimental data but rather to show the effects qualitatively. Also most of the parameters of the model cannot be measured in experiments (for instance, $\lambda_t$, $\beta,\tau_*$ etc.). Hence, for plotting both from the analytical theory and the simulations we use the values which allow to obtain the diffusion properties consistent with experimental observations.

\section{Simulation results and analytical theory for the mean-squared displacement (MSD)}

\subsection{Simulations of MSD}

In the experiments one could measure the profile of the exciton density from the direct PL signal as well as the overall photoluminescence output as a function of time \cite{berghuis2021,Deng2020}. If the original profile of the excitation is flat so will be the exciton density in time. If the beam is Gaussian then the density will evolve in time around the original shape. Then the transport properties can be extracted from the shape of the density of exciton distribution $n(\mathbf{r},t)$\ddag. There are different approaches to the visualisation and analysis of this shape in experiments \cite{berghuis2021,beret2023nonlinear,Rosati2020,Rosati2021}. One option consists in plotting the cross-section of exciton densities normalised by the maximum value of the distribution \cite{berghuis2021}. Alternatively one could consider the full width at half maximum or the mean-squared width of the distribution normalised by the number of excitons as in Refs. \cite{Rosati2020}. De facto both methods account for the fact that in the regime of excitation by a short pulse the number of excitons decreases with time. If one does not account for that and keeps the normalisation over the initial number of excitons then all averages tend to zero.

\footnotetext{\ddag~From these profiles one can indirectly extract the diffusion properties by making model assumptions. For any distribution the full width at half maximum (FWHM) can be always estimated. In the case of non-interacting excitons in the systems without traps the distribution stays Gaussian, the excitons perform a normal diffusion with a diffusion coefficient $D=(\sigma^2(t)-\sigma^2(0))/2t$, where $\sigma^2(t)$ is the variance of the distribution as a function of time. Once the traps with a power-law distribution of escape times are included into the consideration this formula stops to work since the motion becomes subdiffusive for the free and the overall number of excitons and PDF curves become non-Gaussian. The same is true in the case with binary annihilation of excitons. The profile becomes non-Gaussian and, hence, one cannot use the variance or FWHM as a correct measure to derive the diffusion coefficient since this variance does not actually correspond to the mean-squared displacement.
}

\begin{figure}
    \includegraphics[width=1.0\columnwidth]{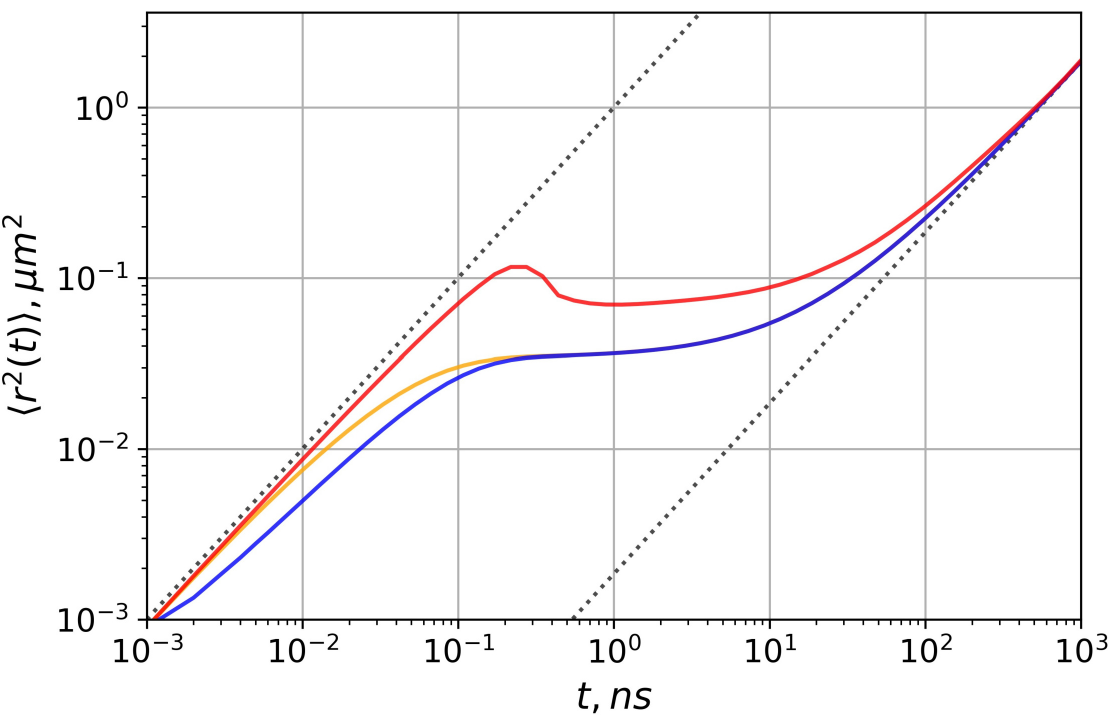}
    \caption{The mean-squared displacement (MSD) obtained in simulation model without recombination and annihilation of excitons, $\alpha=0$, $\beta=0$. Exponential distribution of trapping times has a rate $\lambda_t=0.1\,\mathrm{ns}^{-1}$. The red curve corresponds to MSD of free excitons, the orange curve shows MSD for all excitons and the blue curve illustrates the trapped excitons only. The dotted lines show that the slope at short and at long times is 1 which corresponds to normal diffusion.}
    \label{Fig:NoRecomSim}
\end{figure}

While in experiments one cannot track every exciton it is possible to do that in simulation studies, i.e. one could plot mean-squared displacement (MSD) for a single particle. We start with the analysis of a simple case without recombination or annihilation of excitons. In Fig. \ref{Fig:NoRecomSim} the mean-squared displacement of a single exciton is plotted depending on whether for the cases it either belongs to the free $(\langle r^2(t) \rangle_{f})$ or trapped ($\langle r^2(t) \rangle_{t}$) state or averaged over the whole ensemble ($\langle r^2(t)\rangle$). The MSD is renormalised by the number of surviving excitons as is done in experiments \cite{Rosati2020,Rosati2021},

\begin{equation}\label{MSDdef}
\langle r^2(t)\rangle_{f,t}=\frac{1}{N_{f,t}(t)}\int \mathbf r^2 n_{f,t}(\mathbf{r},t)d\mathbf{r}, 
\end{equation}
where the subscript notations point on whether the quantity is for free ($f$), trapped ($t$) or the overall ensemble of excitons (no subscript). In Fig. \ref{Fig:NoRecomSim} we show the case for which an intermediate negative diffusion is observed. Initially the MSD for free particles grows linearly, i.e. the particles perform a simple random walk. The MSD of trapped excitons also grows linearly at first since the newly trapped excitons are produced from free excitons with the linear MSD. At longer times the MSD evolution changes because the majority of excitons gets trapped. The free exciton population is renewed by the release of trapped excitons. The renewal at this stage is more intense in the centre of the original distribution, because the excitons were trapped there on average earlier. Then the distribution of free excitons gets narrower, and one observes negative diffusion. This phenomenon is essentially transitional and the diffusion again becomes Brownian at long times. We see that to observe the negative diffusion one needs the timescale of getting trapped $1/\lambda_f$ being noticeably shorter than the characteristic release time $1/\lambda_t$ (see also the theoretical considerations below) and yet the diffusion of free excitons should be fast enough. In other cases the observed MSD grows monotonously with a possibility of flattening at the intermediate times which we call a stagnated diffusion (we do not show the simulation results example here since it was done in Refs. \cite{Kurilovich2021,Kurilovich2022}).

\begin{figure}
    \includegraphics[width=1.0\columnwidth]{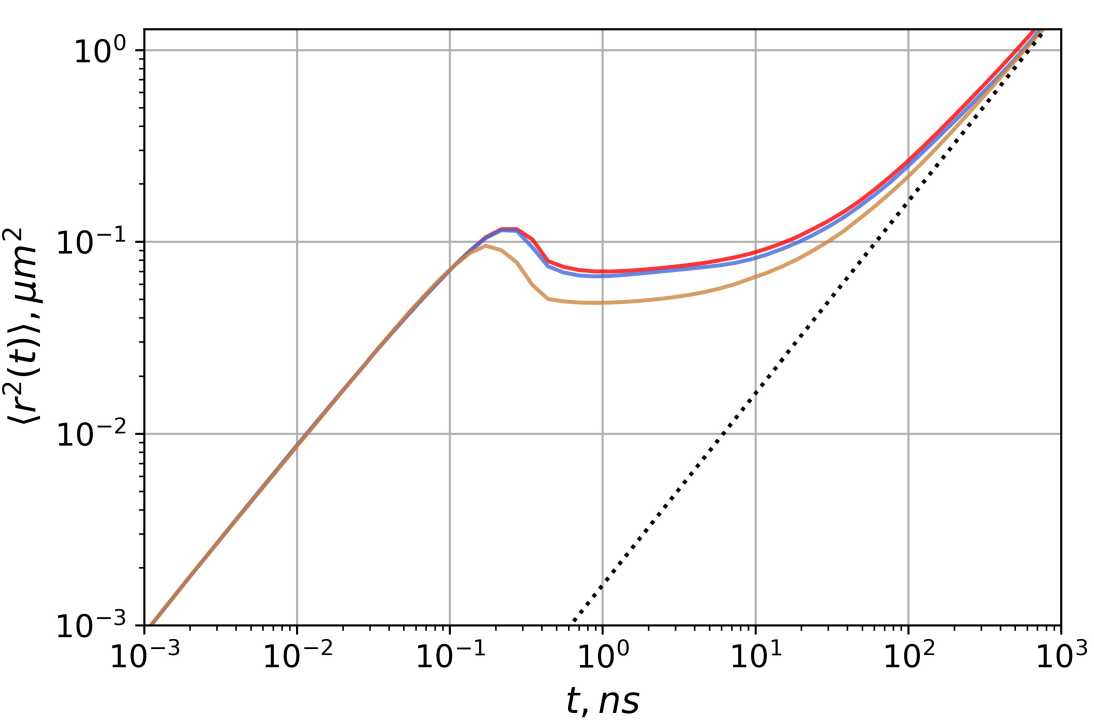}
    \caption{Simulation of the normalised MSD for the free excitons for the cases. Exponential distribution of the trapping times. (a) $\alpha=0,\beta=0$ (red line), (b) $\beta=0,\alpha=1\,\mathrm{ns}^{-1}$ (blue line), (c) $\beta=1\, \mathrm{{\mu m}^{-2}/ns},\alpha=1\,\mathrm{ns}^{-1}$ (brown line). At short times all curves follow simple Brownian diffusion scaling. At long times all curves converge back to the Brownian behaviour (the slope of all of the curves becomes 1).}
    \label{figalphavsnoalpha}
\end{figure}

If we add the linear recombination, i.e. $\alpha\neq0$, the visible effect of the negative diffusion gets enhanced (blue curve in Fig. \ref{figalphavsnoalpha}). If in addition there is an exciton-exciton annihilation (binary reactions) it enhances the observation of negative diffusion even further (brown curve in Fig. \ref{figalphavsnoalpha}). The annihilation effects, however, play the role only at short and intermediate times. At long times the concentration of excitons drops down and the binary reactions become rare.

Interestingly, in our simulation model one of the critical elements is the effect of trap saturation. Once one switches the saturation for the parameters considered, the negative slope of MSD disappears due to an effective drop of trapping rate and the faster effective spread of free excitons.

\begin{figure}[h]
    \includegraphics[width=1.0\columnwidth]{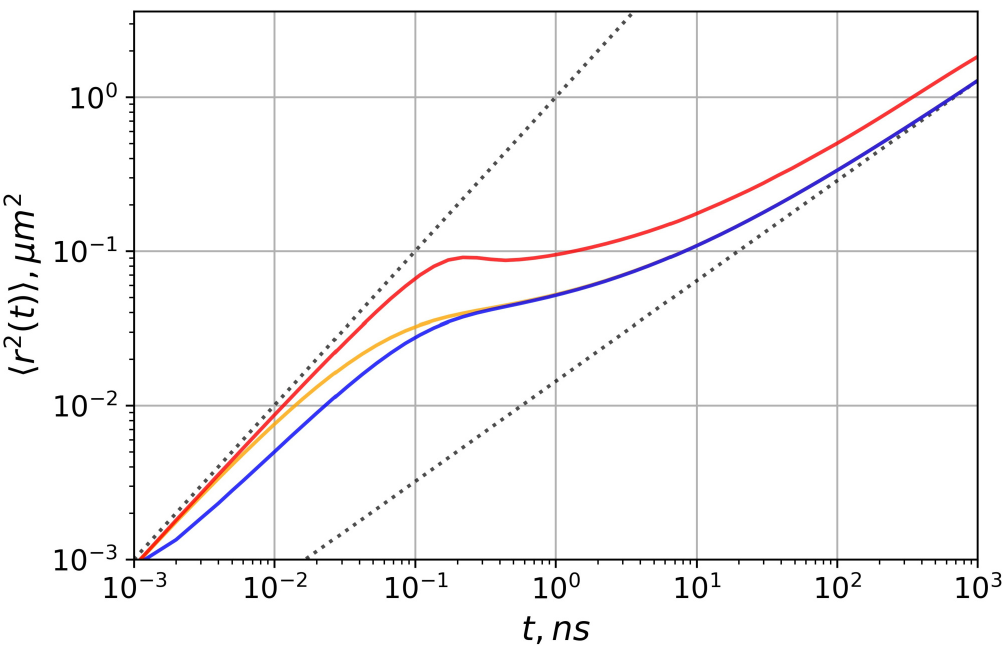}
    \caption{MSD for the Mittag-Leffler distribution ($\mu=0.65,\tau_*=5ns$) of the trapping times could also lead to negative diffusion. The black lines show the asymptotes, with the slope being 1 at short times and $\mu$ at long times (subdiffusion).}
    \label{Fig:MittagLefflerMSD}
\end{figure}

At short times the normal diffusion (the slope is equal to unity) proceeds with the coefficient corresponding to the regular jumps between the neighboring lattice sites. At long times the effective diffusion coefficient is smaller than the free exciton diffusion coefficient by a factor depending on the ratio of trapping/escape rates (also see Eq. (\ref{longlimit}) and Eq. (5.3) in \cite{doerries2022apparent}). If one considers a scale-free distribution of escape times than at long times the slope is subdiffusive, i.e. its value is less than unity, and defined by the corresponding power-law tail of the escape time distribution (Fig. \ref{Fig:MittagLefflerMSD}). The transition region, however, looks similar and exhibits both the case with an effective focussing of the free excitons as well as the monotonous transition to the long-time behaviour.

\subsection{Analytical theory for MSD}
\begin{figure}
    \includegraphics[width=1.0\columnwidth]{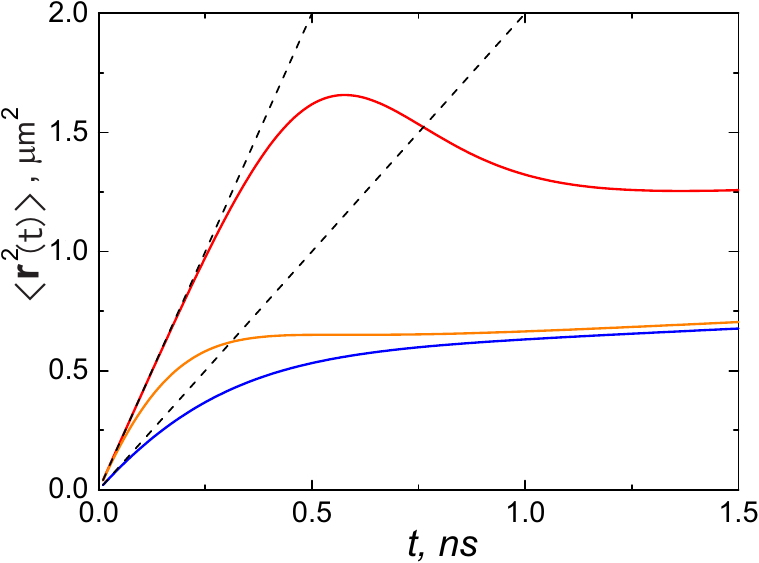}
    \caption{Mean-squared width for free, trapped and all excitons. The continuous lines correspond to free (red), trapped (blue) and total (orange) numbers of excitons and plotted from analytical formulae in Appendix \ref{App:Widthanalyt}. The black dotted straight lines are plotted from linear terms in Eqs. (\ref{analytshorttimeWidthfree})-(\ref{analytshorttimeWidthtotal}), i.e. they read $2Dt$ and $4Dt$. $D=1\,\mu m^2/\mathrm{ns}$. $\alpha=4\, \mathrm{ns^{-1}},\lambda_f=3 \, \mathrm{ns^{-1}},\lambda_t=1/3\, \mathrm{ns^{-1}}$.}
    \label{Fig:WidththeoryLinLin}
\end{figure}

\begin{figure*}
    \includegraphics[width=\linewidth]{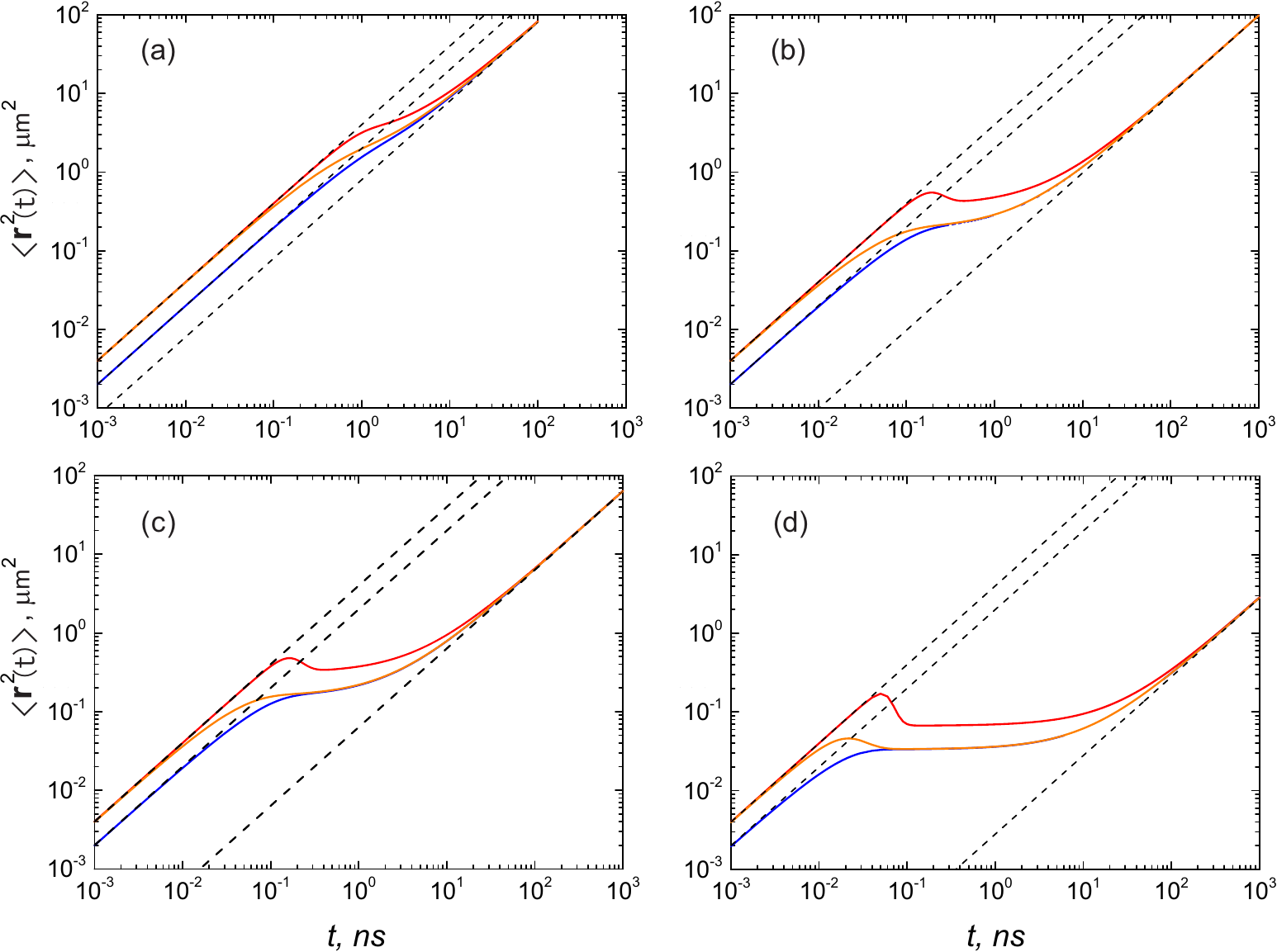}
    
    \caption{Average squared width or MSD for free, trapped and all excitons plotted from analytical expressions in Appendix \ref{App:Widthanalyt}. The continuous lines correspond to free (red), trapped (blue) and overall (orange) numbers of excitons. The black dashed straight lines with the slope 1 are plotted according to Eqs. (\ref{analytshorttimeWidthfree})-(\ref{analytshorttimeWidthtotal}) for the short times' limiting case and for the long times' case from Eq. (\ref{Eq:WidthfreeLongAnalyt}). $D=1\,\mu m^2/\mathrm{ns}$. (a) $\alpha=0,\lambda_f=2\, \mathrm{ns^{-1}},\lambda_t=1/2\, \mathrm{ns^{-1}}$, (b) $\alpha=0,\lambda_f=20\, \mathrm{ns^{-1}},\lambda_t=1/2\, \mathrm{ns^{-1}}$, (c) $\alpha=5\, \mathrm{ns^{-1}},\lambda_f=20\, \mathrm{ns^{-1}},\lambda_t=1/2\, \mathrm{ns^{-1}}$, (d) $\alpha=100\, \mathrm{ns^{-1}},\lambda_f=20\, \mathrm{ns^{-1}},\lambda_t=1/2\, \mathrm{ns^{-1}}$.}
    \label{Fig:Widththeory}
\end{figure*}

In the absence of non-linear term Eqs. (\ref{nf})-(\ref{nt}) can be solved analytically in the inverse Fourier-Laplace space,
\begin{eqnarray}
    &n_f (\mathbf{k}, s) = \frac{N_0}{s +\alpha+ \lambda_f\left[1-\tilde\gamma(s)\right] + D|\mathbf{k}|^2},\label{freedensitysolution}\\
    &n_t (\mathbf{k}, s) =\frac{N_0\lambda_f[1-\tilde\gamma(s)]}{s\left\{s + \alpha+\lambda_f[1-\tilde\gamma(s)] + D|\mathbf{k}|^2\right\}},
    \label{trappeddensitysolution}
\end{eqnarray}
where for the exponential distribution of trapping times the Laplace image of the trapping time distribution $\tilde\gamma(s)=\lambda_t/(s+\lambda_t)$. In the case of the scale free distributions of waiting times in a trap the Eqs. (\ref{freedensitysolution}),(\ref{trappeddensitysolution}) stay valid but the form of $\tilde\gamma(s)$ changes.

The image of the total concentration of all excitons $n(\mathbf{r},t)$ is then
\begin{equation}
    n(\mathbf{k}, s) =\frac{N_0\left(s+\lambda_f[1-\tilde\gamma(s)]\right)}{s\left\{s + \alpha+\lambda_f[1-\tilde\gamma(s)] + D|\mathbf{k}|^2\right\}}.
\end{equation}
For the case of the exponential trapping times one can compute analytically the normalised mean-squared displacement as well as the overall number of mobile and immobile excitons (for the explicit expressions of the latter see the next section). The overall number of free excitons in the case of only linear recombination will also be proportional to the overall PL intensity measured in experiments \cite{Rabouw2016,hinterding2021single}.

As was mentioned above one can measure experimentally the distribution of the exciton density under the assumption that the density of excitons is proportional to the local PL signal. This allows to estimate the squared spatial width of the distribution \cite{Rosati2020}. In the theoretical approach this quantity can be defined for the free/trapped excitons as
\begin{equation}\label{Eq:WidthDef}
\langle  r^2(t) \rangle_{f,t}=\frac{1}{N_{f,t}(t)}\int  \mathbf r^2 n_{f,t}(\mathbf{r},t)d\mathbf{r}=\frac{I_{f,t}(t)}{N_{f,t}(t)},    
\end{equation}
where the integral $I_{f,t}(t)$ is normalised by the corresponding time-dependent number of excitons. This normalisation is necessary since the excitons numbers change and eventually tend to zero due to the recombination. In our theoretical model this is also equivalent to MSD due to the selected initial conditions. The integrals can be computed in the Laplace space similar to the case $\alpha=0$ considered in Ref. \cite{Kurilovich2022},
\begin{eqnarray}\label{Eq:WidthNonFree}
I_{f}(s)=\frac{4DN_0}{(s+\alpha+\lambda_f(1-\tilde\gamma(s)))^2},\\\label{Eq:WidthNonTrapped}
I_{t}(s)=\frac{4DN_0\lambda_f(1-\tilde\gamma(s))}{s(s+\alpha+\lambda_f(1-\tilde\gamma(s)))^2},\\\label{Eq:WidthNonTotal}
I(s)=\frac{4DN_0(s+\lambda_f(1-\tilde\gamma(s))}{s(s+\alpha+\lambda_f(1-\tilde\gamma(s)))^2}.
\end{eqnarray}
One could find the inverse Laplace transform of the latter equations analytically. These are, however, somewhat cumbersome and is shown in Appendix \ref{App:Widthanalyt}.

In the limit of short times, however, the expressions simplify (for the expressions for $N_{t,f}(t)$ see the next section), and one can find the mean-squared width of the distributions (or MSD),
\begin{eqnarray}\label{analytshorttimeWidthfree}
&&\langle r^2(t) \rangle_{f}\simeq4Dt-\frac{2\lambda_f\lambda_t t^3}{3}+...\\\label{analytshorttimeWidthtrapped}
&&\langle r^2(t) \rangle_{t}\simeq2Dt+\frac{1-(\alpha+\lambda_f)/\lambda_t}{3}\lambda_tt^2+...,\\\label{analytshorttimeWidthtotal}
&&\langle r^2(t) \rangle\simeq4Dt-2 \lambda_f t^2+... .
\end{eqnarray}
While all three populations of excitons demonstrate Brownian motion at short times the fastest moving are the free excitons while the effective diffusion coefficient of trapped excitons is twice smaller. The corrections correspond to the tendencies seen in simulations (cf. Fig. \ref{Fig:NoRecomSim},\ref{figalphavsnoalpha}).
At long times all populations of excitons follow the same evolution of the squared spatial width/MSD (for the derivation see Appendix \ref{App:Widthanalyt}),
\begin{equation}\label{Eq:WidthfreeLongAnalyt}
    \langle  r^2(t) \rangle_{f}=\langle r^2(t) \rangle_{t}=\langle r^2(t) \rangle=4DC_1(\alpha,\lambda_f,\lambda_t)t,
\end{equation}
where $C_1$ is a constant which depends on $\alpha,\lambda_f,\lambda_t$ and is defined in Appendix \ref{App:Widthanalyt}. 

In the case $\alpha=0$ all expressions are simplified and allow some qualitative analysis. For the long time limit they read
\begin{equation}\label{longlimit}
    \langle r^2(t) \rangle_{f}=\langle r^2(t) \rangle_{t}=\langle r^2(t) \rangle=\frac{4Dt}{1+\lambda_f/\lambda_t},
\end{equation}
i.e. the long term diffusion coefficient is adjusted by the ratio of trapping and escape rates. One can also see that $\frac{\partial\langle r^2(t) \rangle_{f}}{\partial t}\vert_{t\rightarrow0}>0$ and $\frac{\partial\langle  r^2(t) \rangle_{f}}{\partial t}\vert_{t\rightarrow\infty}>0$ while for $\lambda_f\ll t\ll\lambda_t$ the MSD for the free particles exhibits a plateau regime which we will also call as stagnated diffusion, $\langle r^2(t) \rangle_{f}\approx8D/\lambda_f$. In the same case $\lambda_f\ll t\ll\lambda_t$ for all excitons the MSD reaches plateau at the half value, $\langle r^2(t) \rangle\approx4D/\lambda_f$. The prefactors for the plateau are twice higher than in the one-dimensional MIM model with $\alpha=0$ in Eq. (3.7) from Ref. \cite{doerries2022apparent} since here we consider the two-dimensional case. Also one could check that the derivative of MSD of all excitons is always positive if there is no recombination, i.e. overall population of excitons cannot show negative diffusion unless $\alpha\neq0$. 

For the trapping distribution with the power-law tail  (Mittag-Leffler, for instance,  \cite{Kurilovich2022}) the mean-squared displacement can be obtained by inverting the corresponding formula numerically. The results could produce the similar shape of a "bump" for free excitons and an interval of time with stagnated diffusion depending on the parameters while the long time asymptotic changes to the subdiffusive spread, i.e. $\langle r^2(t)\rangle_{f}=\langle r^2(t) \rangle_{t}=\langle  r^2(t) \rangle=const\, t^\mu,\mu<1$ (not shown) as in simulations in Fig. \ref{Fig:MittagLefflerMSD}.

We first show the dependence of mean squared width or MSD from Eq. (\ref{Eq:WidthDef}) in double linear scale (Fig. \ref{Fig:WidththeoryLinLin}) as is done in Fig. 3 of Ref. \cite{Rosati2020} for the squared spatial width for the system with bright and dark excitons in monolayers of $\mathrm{WS_2}$ dichalcogenide. For the chosen set of parameters one can see the peak and the subsequent decrease of the MSD for mobile excitons. For the trapped and overall populations we observe a monotonous growth. In order to see the scaling as well as to plot longer timescales we move towards the double logarithmic scale in Fig. \ref{Fig:Widththeory}. Here one can see the effect of different parameters on the shape of the MSD curves. When the trapping and the release times are comparable while the recombination is absent ($\alpha=0$, Fig. \ref{Fig:Widththeory}a) MSD for all types of excitons grows monotonously. In Fig. \ref{Fig:Widththeory}b one can see that a substantial increase of a trapping rate $\lambda_f$ causes the formation of a peak on MSD plot and a stagnated diffusion transition region. The presence of recombination (Figs. \ref{Fig:Widththeory}c,d) enhances both effects at the intermediate times. The larger $\alpha$ effects in longer and more pronounced plateau. One could also notice from Fig. \ref{Fig:Widththeory} that the numerical ratio of the MSD plateau values of free excitons to the plateau values for all excitons is $\langle r^2(t) \rangle_f/\langle r^2(t) \rangle\approx2$ even if $\alpha\neq0$.

While the quadratic term removes the possibility of analytical solution of the full set of equations (\ref{nf})-(\ref{nt}), for the homogeneous distribution of the starting positions of excitons the kinetic equations for the exponential trapping times could be written as 
\begin{eqnarray}
&&\frac{dn_f}{dt} =-\alpha n_f - \lambda_fn_f+\lambda_tn_t-\beta n_f^2,\\
&&\frac{dn_t}{dt} =\lambda_fn_f-\lambda_tn_t.
\end{eqnarray}
Unfortunately even this simplification does not allow for the analytical solution of this system. The only way to include the non-linearity analytically then is by treating the binary reactions within the perturbation theory approach. However, the expressions are immensely bulky while provide very little useful information. If plotted the solution with the first-order correction (not shown here) reveals that the binary interactions result in a decrease of the overall number of excitons for every time moment in comparison to the case without them, but no qualitative change occurs.

\section{Number of excitons as a function of time and PL intensity}

\begin{figure}
    \includegraphics[width=1.0\columnwidth]{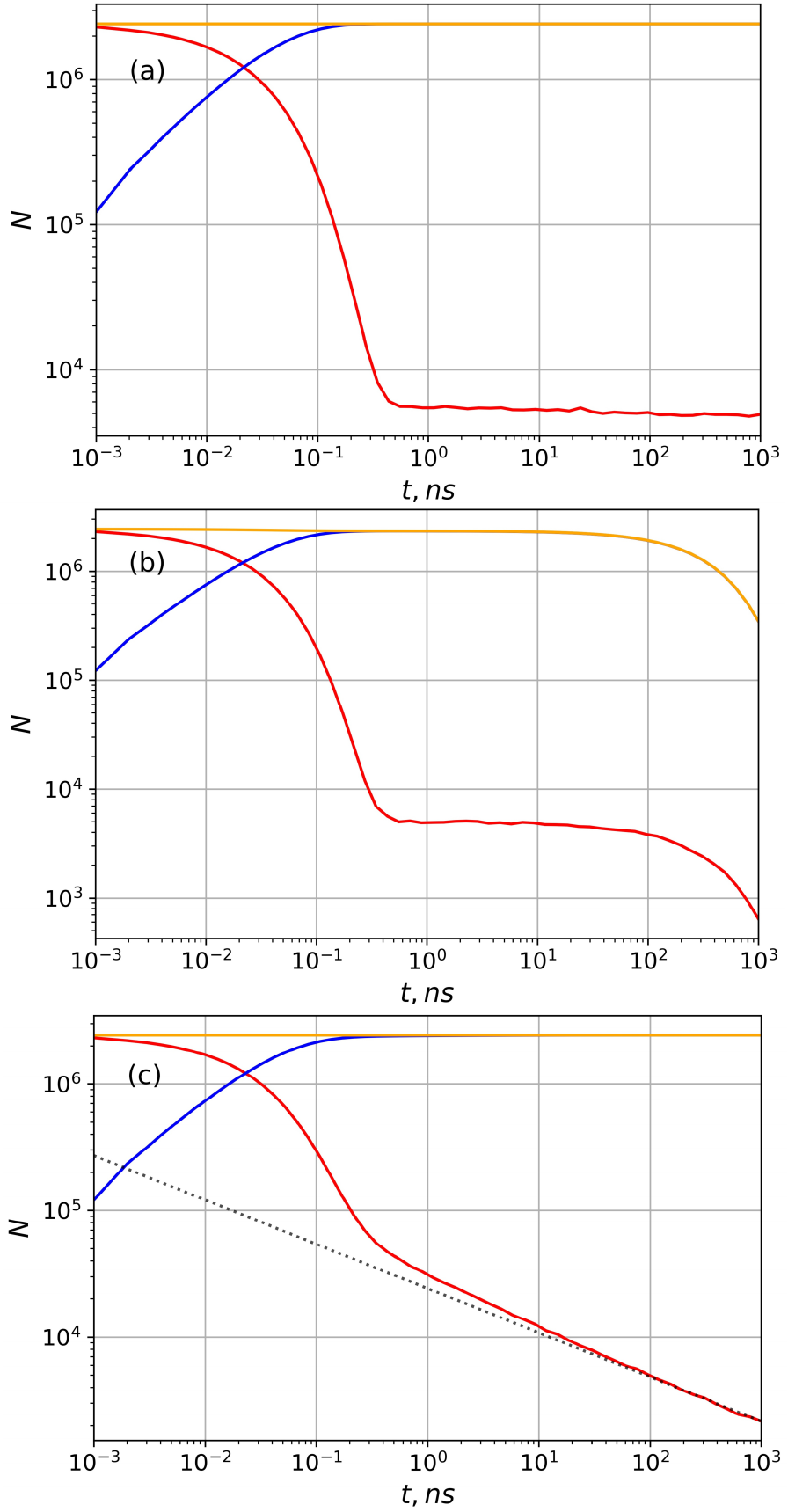}
    \caption{Simulation model results. The number of excitons as a function of time. The three curves correspond to trapped (immobile), free (mobile) particles and both groups together. (a) Exponential trapping, no recombination $\beta=0,\alpha=0$, (b) Exponential trapping with recombination $\beta=0,\alpha=1\, \mathrm{ns^{-1}}$, (c) Mittag-Leffler trapping with the characteristic exponent $\mu=0.65$ and $\tau^*=5\, \mathrm{ns}$ for the distribution, $\beta=0,\alpha=0$.}
    \label{Fig:Nsimul}
\end{figure}

Apart from looking at distributions of excitons and their second moments one could analyse the evolution of the numbers of excitons in different states. The quantity itself is important since the number of free excitons is proportional to the PL intensity \cite{Rabouw2016, hinterding2021single} in the case of a linear decay and close to it when the binary annihilation is taken into the account. In all cases at the initial time moment all excitons are free. Then the numbers are defined by the competition of trapping, recombination and escape of excitons. In Fig. \ref{Fig:Nsimul} we show the typical dependences obtained in our simulation model. The tail for the time dependence of the number of free excitons is defined by the tail of the escape time distribution and the presence or absence of recombination. In the absence of recombination and the exponential trapping the numbers of free/trapped/all excitons tend to constant values (Fig. \ref{Fig:Nsimul}a). The addition of recombination provides a second time scale for the drop numbers of all species of excitons which could be seen in Fig. \ref{Fig:Nsimul}b. If the escape time distribution has a power law asymptotics $\sim\frac{1}{t^{\mu+1}}$, (Fig. \ref{Fig:Nsimul}c) then the number of free excitons decays asymptotically as $\sim\frac{1}{t^{1-\mu}}$ in the case without recombination (the dotted black line in Fig. \ref{Fig:Nsimul}c has a slope $0.65-1=-0.35$) and as $\sim\frac{1}{t^{\mu+1}}$ in the case with recombination (cf. \cite{Kurilovich2020},\cite{Kurilovich2022}).


\begin{figure*}
    \includegraphics[width=\linewidth]{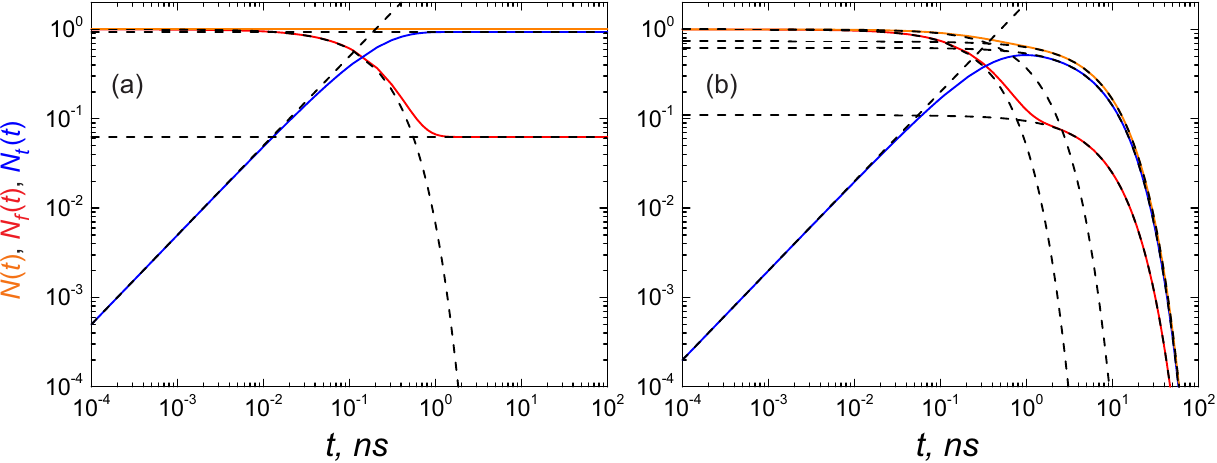}

    \caption{Theoretical results for the number of free, trapped and overall for the linear recombination and without it in units of $N_0$. The continuous lines correspond to free (red), trapped (blue) and overall (orange) numbers of excitons and plotted from analytical formulae (\ref{FullExpressionsNumbers}). The black dotted lines correspond to the short and long time limiting behaviour according to Eqs. (\ref{analytshorttimeNfree})-(\ref{analytshorttimeNtotal}) and (\ref{NumbersLimiting}), correspondingly. The diffusion coefficient $D=1 \mu m^2/ns$. (a) $\alpha=0,\lambda_f=5 \, \mathrm{ns^{-1}}, \lambda_t=1/3 \, \mathrm{ns^{-1}}$, (b) $\alpha=1\, \mathrm{ns^{-1}},\lambda_f=2\, \mathrm{ns^{-1}}, \lambda_t=0.5\, \mathrm{ns^{-1}}$.}
    \label{Fig:Ntheory}
\end{figure*}

The overall number of the corresponding subspecies of excitons can be obtained from Eqs. (\ref{nf})-(\ref{nt}) just by considering $\mathbf{k}=\mathbf{0}$ since 
\begin{eqnarray}
\tilde N_{f,t}(s) = \int \tilde n_{f,t}(\mathbf{r}, s)d\mathbf{r} = \hat{\tilde{n}}_{f,t}(\mathbf{k}=\mathbf{0}, s),
\end{eqnarray}
i.e.
\begin{eqnarray}
    &\tilde N_f (s) = \frac{N_0}{s +\alpha+ \lambda_f\left[1-\tilde\gamma(s)\right]},\label{freenumbersolution}\\
    &\tilde N_t (s) =\frac{N_0\lambda_f[1-\tilde\gamma(s)]}{s\left\{s + \alpha+\lambda_f[1-\tilde\gamma(s)]\right\}},
    \label{trappednumbersolution}
\end{eqnarray}
While for a general form of $\tilde\gamma(s)$ one cannot analytically invert Eqs. (\ref{freenumbersolution})-(\ref{trappednumbersolution}) it is possible for an exponential distribution of trapping times (the full analytical expressions can be found in the Appendix \ref{App:Nanalyt}). The asymptotics look simpler than general expressions. For $t\rightarrow0$, 
\begin{eqnarray}\label{analytshorttimeNfree}
&&N_f(t)\simeq N_0(1-(\alpha+\lambda_f)t+...),\\ \label{analytshorttimeNtrapped} &&N_t(t)\simeq N_0\lambda_f t,\\\label{analytshorttimeNtotal} &&N(t)=N_0(1-\alpha t).
\end{eqnarray}
The case $t\rightarrow\infty$ can be obtained as it corresponds to $s\rightarrow0$. Then the inverse Laplace transform can be easily performed and the resulting formulae (\ref{NumbersLimiting}) are shown in Appendix \ref{App:Nanalyt}.

The results obtained from the above equations are plotted in Fig. \ref{Fig:Ntheory}. Two distinct cases appear. The first one (Fig. \ref{Fig:Ntheory} (a)) is for exponential distribution of the trapping times in the absence of recombination. It corresponds to the case shown in Fig. \ref{Fig:Nsimul}a. An initial exponential decay of the number of free excitons due to the trapping and the linear growth of the number of trapped ones shapes into the equilibrium between these two states and the numbers become constant (the analytical limiting behaviour is shown by black dotted lines). The second scenario occurs if the recombination is possible (Fig. \ref{Fig:Ntheory} (b)). Then at short times the exponential decay for free excitons happens both due to the recombination and the trapping, while the trapped population growths linear as in the other case. The most distinctive change happens at long times where numbers of all subsets of excitons drop exponentially with exponents shown in Appendix \ref{App:Nanalyt}. We see the full consistency of the analytical theory with our simulation model (cf. Fig. \ref{Fig:Nsimul} and Fig. \ref{Fig:Ntheory}).

The power-law tail distribution of trapping times changes the behaviour in a qualitative way. For the number of free excitons the analytical limiting cases were considered in Ref. \cite{Kurilovich2020} and solutions for $\alpha=0$ were studied in \cite{Kurilovich2022}. The decrease is observed even without active recombination, i.e. all excitons eventually move into the trapped state.

\section{Conclusions and Discussion}

Deviations from Brownian diffusive behaviour are manifested in plethora of various physical, biological and other systems \cite{AndreyPCCP2014}. Experimental results for diffusion of excitons in low-dimensional semiconductors often show anomalies in photoluminescence as well as in the shape of the exciton density \cite{Rabouw2016,Seitz2020}. In some of the cases the results point out to the focussing of the density profile which corresponds to the negative diffusion.  In order to explain the phenomenon of negative diffusion we propose and solve simulation and analytical mobile-immobile models. The phenomenon appears due to the interplay of two species of particles, mobile and immobile ones. If one of the species is immobile (or slow) then the fast trapping and the slow release could cause narrowing of the exciton density or decrease of observable MSD at intermediate timescales. The short time behaviour will be Brownian while the long time behaviour will be defined by the trapping time distribution. If the latter is exponential (or with a finite first moment) then the long time diffusion is Brownian as well but with an effective diffusion coefficient defined by trapping. In the case of scale-free waiting times in a trap the subdiffusion is observed. The behaviour of MSD at intermediate timescales will often exhibit a stagnated diffusion. As of now the models and systems which can produce the stagnated diffusion are still a few. Among the others it is worth to mention the MSD of non-normalisable quasi-equilibrium states under fractional dynamics \cite{Eli2023} and probe particle tracking experiments in micellar solutions with subdiffusive dynamics \cite{bellour2002brownian,galvan2008diffusing,jeon2013anomalous} and in the cytosol of bacteria, yeast, and human cells \cite{aaberg2021glass,corci2023extending}.

In contrast to Refs. \cite{berghuis2021,wietek2023non} our theoretical mobile-immobile model explains the possibility of negative diffusion effect even in the absense of recombination or annihilation of excitons. Both the exact analytical solutions and simulation results show that stagnated diffusion regime grows longer and the negative diffusion enhances once recombination is taken into account. Moreover, the annihilation through binary exciton-exciton interactions strengthens the effect further.

\section*{Conflicts of interest}
There are no conflicts to declare.

\section*{Acknowledgements}
V.N.M. acknowledges the support from the Interdisciplinary Scientific and Educational School of Moscow State University ‘‘Photonic and Quantum technologies. Digital medicine’’. A.V.C. acknowledges support from the Polish National Agency for Academic Exchange (NAWA).

\appendix 

\section{Trapping time distributions used in simulations and theory}\label{AppendixMittagLeffler}

The two-parametric Mittag-Leffler function is defined by the power series~\cite{gorenflo2020}
\begin{equation}\label{MittagLefflerEq}
    E_{\nu,\kappa}(z)=\sum_{k=0}^{\infty}\frac{z^k}{\Gamma(\nu k+\kappa)}, 
\end{equation}
where $\Gamma(x)$ is the gamma function. The case $\nu=\kappa=1$ corresponds to the exponential function. In analytical calculations we use the following form of the trapping time distribution, 
\begin{align}
    \gamma(t) = \frac{t^{\mu-1}}{{\tau_*}^\mu}E_{\mu, \mu}\left[-\left(\frac{t}{\tau_*}\right)^\mu\right];\quad 0<\mu\leq 1 .
    \label{eq:mldist}
\end{align}
This function is positively defined, normalised and has a simple Laplace transform given by
\begin{align}
    \tilde\gamma(s) = \frac{1}{1 + (\tau_* s)^\mu}.
    \label{eq:mllaplacedist}
\end{align}
At small $s$ one gets $\tilde\gamma(s) \approx 1 - (\tau_* s)^\mu$ which corresponds to slowly decaying power-law asymptotics $\gamma(t)\sim\tau_*^\mu/t^{1+\mu}$ in real time domain. To sample random variables from the Mittag-Leffler distribution with $\nu=\kappa=\mu$, we used the method suggested in~\cite{Haubold2011}.

\section{Analytical formulas for the mean-squared displacement of excitons for the exponential trapping times}\label{App:Widthanalyt}

The width of the distribution of all considered subsets of excitons can be computed from Eq.  (\ref{Eq:WidthDef}). It looks rather cumbersome, hence we put here the expressions for $I_f(t),I_t(t),I(t)$ from which one gets the formulas according to the definition (\ref{Eq:WidthDef}) by simple division on the corresponding expressions (\ref{FullExpressionsNumbers}).

The non-normalised MSDs are
\begin{widetext}
\begin{eqnarray}\nonumber
&&I_f(t)=\frac{2\exp\left(-(\lambda_t+\alpha+\lambda_f)t/2\right)}{A^{3/2}}\left[\exp\left(\frac{\sqrt A\lambda_tt}{2}\right)\left\{4\lambda_f/\lambda_t^2+t(1+\sqrt A+1/\lambda_t(-3\alpha+\lambda_f +(3\alpha-\lambda_f)(\alpha+\lambda_f)/\lambda_t\right.\right.\\\nonumber&&\left.\left.-(\alpha+\lambda_f)^3/\lambda_t^2+\sqrt A(-2\alpha+(\alpha+\lambda_f)^2/\lambda_t)))\right\}+\right.\\\nonumber&&\left.\exp\left(-\frac{\sqrt A\lambda_tt}{2}\right)\left\{-4\lambda_f/\lambda_t^2+t(-1+\sqrt A+1/\lambda_t(3\alpha-\lambda_f -(3\alpha-\lambda_f)(\alpha+\lambda_f)/\lambda_t\right.\right.\\&&\left.\left.+(\alpha+\lambda_f)^3/\lambda_t^2+\sqrt A(-2\alpha+(\alpha+\lambda_f)^2/\lambda_t)))\right\}\right],\\\nonumber
&&I_t(t)=\frac{2\lambda_f\exp\left(-(\lambda_t+\alpha+\lambda_f)t/2\right)}{A^{3/2}\lambda_t}\left[\exp\left(\frac{\sqrt A\lambda_tt}{2}\right)\left\{2(-1+(\alpha+\lambda_f)/\lambda_t)/\lambda_t+t(1+\sqrt A-\right.\right.\\\nonumber&&\left.\left.1/\lambda_t(2(\alpha-\lambda_f)-(\alpha+\lambda_f)^2/\lambda_t+(\alpha+\lambda_f)\sqrt A))\right\}+\right.\\\nonumber&&\left.\exp\left(-\frac{\sqrt A\lambda_tt}{2}\right)\left\{2(1-(\alpha+\lambda_f)/\lambda_t)/\lambda_t+t(-1+\sqrt A-1/\lambda_t(-2(\alpha-\lambda_f)+(\alpha+\lambda_f)^2/\lambda_t+(\alpha+\lambda_f)\sqrt A))\right\}\right],\\
&&I(t)=I_f(t)+I_t(t),
\end{eqnarray}
\end{widetext}
where $A=-4\alpha/\lambda_t + (1 +(\alpha+ \lambda_f)/\lambda_t)^2$.
Using the above expressions, the results from Appendix \ref{App:Nanalyt} and Eq. (\ref{Eq:WidthDef}) one can derive the limiting cases for the evolution of the mean squared width for the exciton position distributions. At short times the results can be found in Eqs. (\ref{analytshorttimeWidthfree})-(\ref{analytshorttimeWidthtotal}).

In the long time limit $t\rightarrow\infty$ it leads to the following expression for the mean-squared width of the distributions of free, trapped and all excitons,
\begin{eqnarray}\nonumber
 \langle r^2(t) \rangle_{f}=\langle r^2(t) \rangle_{t}=\\\langle r^2(t) \rangle=4DC_1(\alpha,\lambda_f,\lambda_t)t,
\end{eqnarray}
where
\begin{equation}
C_1(\alpha,\lambda_f,\lambda_t)=\frac{\sqrt{A}+(1-(\alpha+\lambda_f)/\lambda_t)}{2\sqrt{A}}
\end{equation}
and $A$ was defined earlier.

\section{Analytical formulas for the numbers of excitons for the exponential trapping times}\label{App:Nanalyt}

In the case of an exponential waiting times $\tilde\gamma(s)=\lambda_t/(s+\lambda_t)$. Then Eqs. (\ref{freenumbersolution})-(\ref{trappednumbersolution}) can be inverted back from Laplace space to the real time analytically,
\begin{widetext} 
\begin{eqnarray}\nonumber
N_f(t)&=&\frac{N_0 (-1+(\alpha+\lambda_f)/\lambda_t+\sqrt A)}{2\sqrt A}\exp\left(-\frac{1+(\alpha+\lambda_f)/\lambda_t+\sqrt A}{2}\lambda_tt\right)\\&+&\frac{N_0 (1-(\alpha+\lambda_f)/\lambda_t+\sqrt A)}{2\sqrt A}\exp\left(-\frac{1+(\alpha+\lambda_f)/\lambda_t-\sqrt A}{2}\lambda_tt\right),\\
N_t(t)&=&\frac{N_0\lambda_f}{\lambda_t\sqrt A}\exp\left(-\frac{\lambda_t+\alpha+\lambda_f}{2}t\right)\left[\exp\left(\frac{\sqrt A\lambda_tt}{2}\right)-\exp\left(-\frac{\sqrt A\lambda_tt}{2}\right)\right],\\\nonumber
N(t)=N_f(t)+N_t(t)&=&\frac{N_0 (-1+(\alpha-\lambda_f)/\lambda_t+\sqrt A)}{2\sqrt A}\exp\left(-\frac{1+(\alpha+\lambda_f)/\lambda_t+\sqrt A}{2}\lambda_tt\right)\\&+&\frac{N_0 (1-(\alpha-\lambda_f)/\lambda_t+\sqrt A)}{2\sqrt A}\exp\left(-\frac{1+(\alpha+\lambda_f)/\lambda_t-\sqrt A}{2}\lambda_tt\right),
\end{eqnarray}
\end{widetext} 
For $t\rightarrow0$, $N_f(t)\simeq N_0(1-(\alpha+\lambda_f)t+...),  N_t(t)\simeq \lambda_f t, N(t)=N_0(1-\alpha t)$. For $\lambda_tt\gg1$ ($s\rightarrow0$)
\begin{widetext} 
\begin{eqnarray}\nonumber
&&N_f(t)\simeq\frac{N_0 (1-(\alpha+\lambda_f)/\lambda_t+\sqrt A)}{2\sqrt A}\exp\left(-\frac{1+(\alpha+\lambda_f)/\lambda_t-\sqrt A}{2}\lambda_tt\right),\\\nonumber
&&N_t(t)\simeq\frac{N_0 \lambda_f}{\lambda_t\sqrt A}\exp\left(-\frac{1+(\alpha+\lambda_f)/\lambda_t-\sqrt A}{2}\lambda_tt\right),\\
&&N(t)\simeq\frac{N_0 (1-(\alpha-\lambda_f)/\lambda_t+\sqrt A)}{2\sqrt A}\exp\left(-\frac{1+(\alpha+\lambda_f)/\lambda_t-\sqrt A}{2}\lambda_tt\right).\label{FullExpressionsNumbers}
\end{eqnarray}
\end{widetext} 

For $\alpha=0$ the equations (\ref{FullExpressionsNumbers}) are substantially simplified and read

\begin{eqnarray}\nonumber
&&N_{f}(t) = N_0\frac{\lambda_t + \lambda_f \exp\left(-t (\lambda_t + \lambda_f)\right)}{\lambda_t + \lambda_f},\\
&&N_{t}(t) = N_0\frac{\lambda_f\left(1 - \exp\left(-t (\lambda_t + \lambda_f)\right)\right)}{\lambda_t + \lambda_f}.\label{NumbersLimiting}
\end{eqnarray}

Naturally,  $N(t)=N_f(t)+N_t(t)=N_0=const$ for $\alpha=0$.


\providecommand*{\mcitethebibliography}{\thebibliography}
\csname @ifundefined\endcsname{endmcitethebibliography}
{\let\endmcitethebibliography\endthebibliography}{}



\begin{mcitethebibliography}{97}
\providecommand*{\natexlab}[1]{#1}
\providecommand*{\mciteSetBstSublistMode}[1]{}
\providecommand*{\mciteSetBstMaxWidthForm}[2]{}
\providecommand*{\mciteBstWouldAddEndPuncttrue}
  {\def\EndOfBibitem{\unskip.}}
\providecommand*{\mciteBstWouldAddEndPunctfalse}
  {\let\EndOfBibitem\relax}
\providecommand*{\mciteSetBstMidEndSepPunct}[3]{}
\providecommand*{\mciteSetBstSublistLabelBeginEnd}[3]{}
\providecommand*{\EndOfBibitem}{}
\mciteSetBstSublistMode{f}
\mciteSetBstMaxWidthForm{subitem}
{(\emph{\alph{mcitesubitemcount}})}
\mciteSetBstSublistLabelBeginEnd{\mcitemaxwidthsubitemform\space}
{\relax}{\relax}

\bibitem[Pope \emph{et~al.}(1999)Pope, Swenberg, and Swenberg]{pope1999}
M.~Pope, C.~Swenberg and P.~Swenberg, \emph{Electronic Processes in Organic
  Crystals and Polymers}, Oxford University Press, 1999\relax
\mciteBstWouldAddEndPuncttrue
\mciteSetBstMidEndSepPunct{\mcitedefaultmidpunct}
{\mcitedefaultendpunct}{\mcitedefaultseppunct}\relax
\EndOfBibitem
\bibitem[Colby \emph{et~al.}(2010)Colby, Burdett, Frisbee, Zhu, Dillon, and
  Bardeen]{colby2010}
K.~A. Colby, J.~J. Burdett, R.~F. Frisbee, L.~Zhu, R.~J. Dillon and C.~J.
  Bardeen, \emph{The Journal of Physical Chemistry A}, 2010, \textbf{114},
  3471--3482\relax
\mciteBstWouldAddEndPuncttrue
\mciteSetBstMidEndSepPunct{\mcitedefaultmidpunct}
{\mcitedefaultendpunct}{\mcitedefaultseppunct}\relax
\EndOfBibitem
\bibitem[Berghuis \emph{et~al.}(2021)Berghuis, Raziman, Halpin, Wang, Curto,
  and Rivas]{berghuis2021}
A.~M. Berghuis, T.~Raziman, A.~Halpin, S.~Wang, A.~G. Curto and J.~G. Rivas,
  \emph{The Journal of Physical Chemistry Letters}, 2021, \textbf{12},
  1360--1366\relax
\mciteBstWouldAddEndPuncttrue
\mciteSetBstMidEndSepPunct{\mcitedefaultmidpunct}
{\mcitedefaultendpunct}{\mcitedefaultseppunct}\relax
\EndOfBibitem
\bibitem[Hadziioannou and van Hutten(2000)]{hadziioannou2000}
G.~Hadziioannou and P.~van Hutten, \emph{Semiconducting Polymers: Chemistry,
  Physics and Engineering}, Wiley, 2000\relax
\mciteBstWouldAddEndPuncttrue
\mciteSetBstMidEndSepPunct{\mcitedefaultmidpunct}
{\mcitedefaultendpunct}{\mcitedefaultseppunct}\relax
\EndOfBibitem
\bibitem[Bolinger \emph{et~al.}(2011)Bolinger, Traub, Adachi, and
  Barbara]{bolinger2011}
J.~C. Bolinger, M.~C. Traub, T.~Adachi and P.~F. Barbara, \emph{Science}, 2011,
  \textbf{331}, 565--567\relax
\mciteBstWouldAddEndPuncttrue
\mciteSetBstMidEndSepPunct{\mcitedefaultmidpunct}
{\mcitedefaultendpunct}{\mcitedefaultseppunct}\relax
\EndOfBibitem
\bibitem[Kim \emph{et~al.}(2007)Kim, Lee, Coates, Moses, Nguyen, Dante, and
  Heeger]{kim2007}
J.~Y. Kim, K.~Lee, N.~E. Coates, D.~Moses, T.-Q. Nguyen, M.~Dante and A.~J.
  Heeger, \emph{Science}, 2007, \textbf{317}, 222--225\relax
\mciteBstWouldAddEndPuncttrue
\mciteSetBstMidEndSepPunct{\mcitedefaultmidpunct}
{\mcitedefaultendpunct}{\mcitedefaultseppunct}\relax
\EndOfBibitem
\bibitem[Ivchenko(2005)]{ivchenko2005}
E.~Ivchenko, \emph{Optical Spectroscopy of Semiconductor Nanostructures}, Alpha
  Science, 2005\relax
\mciteBstWouldAddEndPuncttrue
\mciteSetBstMidEndSepPunct{\mcitedefaultmidpunct}
{\mcitedefaultendpunct}{\mcitedefaultseppunct}\relax
\EndOfBibitem
\bibitem[Takagahara(2003)]{takagahara2003}
T.~Takagahara, \emph{Quantum Coherence Correlation and Decoherence in
  Semiconductor Nanostructures}, Elsevier Science, 2003\relax
\mciteBstWouldAddEndPuncttrue
\mciteSetBstMidEndSepPunct{\mcitedefaultmidpunct}
{\mcitedefaultendpunct}{\mcitedefaultseppunct}\relax
\EndOfBibitem
\bibitem[Scholes and Rumbles(2006)]{scholes2006}
G.~Scholes and G.~Rumbles, \emph{Excitons in nanoscale systems}, 2006,
  \textbf{5}, 683--697\relax
\mciteBstWouldAddEndPuncttrue
\mciteSetBstMidEndSepPunct{\mcitedefaultmidpunct}
{\mcitedefaultendpunct}{\mcitedefaultseppunct}\relax
\EndOfBibitem
\bibitem[Grim \emph{et~al.}(2014)Grim, Christodoulou, Di~Stasio, Krahne,
  Cingolani, Manna, and Moreels]{grim2014}
J.~Q. Grim, S.~Christodoulou, F.~Di~Stasio, R.~Krahne, R.~Cingolani, L.~Manna
  and I.~Moreels, \emph{Nature nanotechnology}, 2014, \textbf{9},
  891--895\relax
\mciteBstWouldAddEndPuncttrue
\mciteSetBstMidEndSepPunct{\mcitedefaultmidpunct}
{\mcitedefaultendpunct}{\mcitedefaultseppunct}\relax
\EndOfBibitem
\bibitem[Diroll \emph{et~al.}(2023)Diroll, Guzelturk, Po, Dabard, Fu, Makke,
  Lhuillier, and Ithurria]{nanoplateletschemrev2023}
B.~T. Diroll, B.~Guzelturk, H.~Po, C.~Dabard, N.~Fu, L.~Makke, E.~Lhuillier and
  S.~Ithurria, \emph{Chemical Reviews}, 2023, \textbf{0}, null\relax
\mciteBstWouldAddEndPuncttrue
\mciteSetBstMidEndSepPunct{\mcitedefaultmidpunct}
{\mcitedefaultendpunct}{\mcitedefaultseppunct}\relax
\EndOfBibitem
\bibitem[Klimov \emph{et~al.}(2000)Klimov, Mikhailovsky, Xu, Malko,
  Hollingsworth, Leatherdale, Eisler, and Bawendi]{klimov2000}
V.~Klimov, A.~Mikhailovsky, S.~Xu, A.~Malko, J.~Hollingsworth, a.~C.
  Leatherdale, H.-J. Eisler and M.~Bawendi, \emph{science}, 2000, \textbf{290},
  314--317\relax
\mciteBstWouldAddEndPuncttrue
\mciteSetBstMidEndSepPunct{\mcitedefaultmidpunct}
{\mcitedefaultendpunct}{\mcitedefaultseppunct}\relax
\EndOfBibitem
\bibitem[Ithurria \emph{et~al.}(2011)Ithurria, Tessier, Mahler, Lobo,
  Dubertret, and Efros]{ithurria2011}
S.~Ithurria, M.~Tessier, B.~Mahler, R.~Lobo, B.~Dubertret and A.~L. Efros,
  \emph{Nature materials}, 2011, \textbf{10}, 936--941\relax
\mciteBstWouldAddEndPuncttrue
\mciteSetBstMidEndSepPunct{\mcitedefaultmidpunct}
{\mcitedefaultendpunct}{\mcitedefaultseppunct}\relax
\EndOfBibitem
\bibitem[Smirnov \emph{et~al.}(2019)Smirnov, Golinskaya, Kotin, Dorofeev,
  Palyulin, Mantsevich, and Dneprovskii]{smirnov2019}
A.~Smirnov, A.~Golinskaya, P.~Kotin, S.~Dorofeev, V.~Palyulin, V.~Mantsevich
  and V.~Dneprovskii, \emph{Journal of Luminescence}, 2019, \textbf{213},
  29--35\relax
\mciteBstWouldAddEndPuncttrue
\mciteSetBstMidEndSepPunct{\mcitedefaultmidpunct}
{\mcitedefaultendpunct}{\mcitedefaultseppunct}\relax
\EndOfBibitem
\bibitem[Smirnov \emph{et~al.}(2019)Smirnov, Golinskaya, Kotin, Dorofeev,
  Zharkova, Palyulin, Mantsevich, and Dneprovskii]{smirnov2019_1}
A.~M. Smirnov, A.~D. Golinskaya, P.~A. Kotin, S.~G. Dorofeev, E.~V. Zharkova,
  V.~V. Palyulin, V.~N. Mantsevich and V.~S. Dneprovskii, \emph{The Journal of
  Physical Chemistry C}, 2019, \textbf{123}, 27986--27992\relax
\mciteBstWouldAddEndPuncttrue
\mciteSetBstMidEndSepPunct{\mcitedefaultmidpunct}
{\mcitedefaultendpunct}{\mcitedefaultseppunct}\relax
\EndOfBibitem
\bibitem[Rabouw \emph{et~al.}(2016)Rabouw, van~der Bok, Spinicelli, Mahler,
  Nasilowski, Pedetti, Dubertret, and Vanmaekelbergh]{Rabouw2016}
F.~T. Rabouw, J.~C. van~der Bok, P.~Spinicelli, B.~Mahler, M.~Nasilowski,
  S.~Pedetti, B.~Dubertret and D.~Vanmaekelbergh, \emph{Nano letters}, 2016,
  \textbf{16}, 2047--2053\relax
\mciteBstWouldAddEndPuncttrue
\mciteSetBstMidEndSepPunct{\mcitedefaultmidpunct}
{\mcitedefaultendpunct}{\mcitedefaultseppunct}\relax
\EndOfBibitem
\bibitem[Shornikova \emph{et~al.}(2018)Shornikova, Biadala, Yakovlev, Sapega,
  Kusrayev, Mitioglu, Ballottin, Christianen, Belykh,
  Kochiev,\emph{et~al.}]{Shornikova2018}
E.~V. Shornikova, L.~Biadala, D.~R. Yakovlev, V.~F. Sapega, Y.~G. Kusrayev,
  A.~A. Mitioglu, M.~V. Ballottin, P.~C. Christianen, V.~V. Belykh, M.~V.
  Kochiev \emph{et~al.}, \emph{Nanoscale}, 2018, \textbf{10}, 646--656\relax
\mciteBstWouldAddEndPuncttrue
\mciteSetBstMidEndSepPunct{\mcitedefaultmidpunct}
{\mcitedefaultendpunct}{\mcitedefaultseppunct}\relax
\EndOfBibitem
\bibitem[Olutas \emph{et~al.}(2015)Olutas, Guzelturk, Kelestemur, Yeltik,
  Delikanli, and Demir]{Olutas2015}
M.~Olutas, B.~Guzelturk, Y.~Kelestemur, A.~Yeltik, S.~Delikanli and H.~V.
  Demir, \emph{Acs Nano}, 2015, \textbf{9}, 5041--5050\relax
\mciteBstWouldAddEndPuncttrue
\mciteSetBstMidEndSepPunct{\mcitedefaultmidpunct}
{\mcitedefaultendpunct}{\mcitedefaultseppunct}\relax
\EndOfBibitem
\bibitem[Brumberg \emph{et~al.}(2019)Brumberg, Harvey, Philbin, Diroll, Lee,
  Crooker, Wasielewski, Rabani, and Schaller]{Brumberg2019}
A.~Brumberg, S.~M. Harvey, J.~P. Philbin, B.~T. Diroll, B.~Lee, S.~A. Crooker,
  M.~R. Wasielewski, E.~Rabani and R.~D. Schaller, \emph{ACS nano}, 2019,
  \textbf{13}, 8589--8596\relax
\mciteBstWouldAddEndPuncttrue
\mciteSetBstMidEndSepPunct{\mcitedefaultmidpunct}
{\mcitedefaultendpunct}{\mcitedefaultseppunct}\relax
\EndOfBibitem
\bibitem[Biadala \emph{et~al.}(2014)Biadala, Liu, Tessier, Yakovlev, Dubertret,
  and Bayer]{Biadala2014}
L.~Biadala, F.~Liu, M.~D. Tessier, D.~R. Yakovlev, B.~Dubertret and M.~Bayer,
  \emph{Nano letters}, 2014, \textbf{14}, 1134--1139\relax
\mciteBstWouldAddEndPuncttrue
\mciteSetBstMidEndSepPunct{\mcitedefaultmidpunct}
{\mcitedefaultendpunct}{\mcitedefaultseppunct}\relax
\EndOfBibitem
\bibitem[Meerbach \emph{et~al.}(2019)Meerbach, Tietze, Voigt, Sayevich,
  Dzhagan, Erwin, Dang, Selyshchev, Schneider,
  Zahn,\emph{et~al.}]{Meerbach2019}
C.~Meerbach, R.~Tietze, S.~Voigt, V.~Sayevich, V.~M. Dzhagan, S.~C. Erwin,
  Z.~Dang, O.~Selyshchev, K.~Schneider, D.~R. Zahn \emph{et~al.},
  \emph{Advanced Optical Materials}, 2019, \textbf{7}, 1801478\relax
\mciteBstWouldAddEndPuncttrue
\mciteSetBstMidEndSepPunct{\mcitedefaultmidpunct}
{\mcitedefaultendpunct}{\mcitedefaultseppunct}\relax
\EndOfBibitem
\bibitem[Mak \emph{et~al.}(2010)Mak, Lee, Hone, Shan, and Heinz]{Mak2010}
K.~F. Mak, C.~Lee, J.~Hone, J.~Shan and T.~F. Heinz, \emph{Physical review
  letters}, 2010, \textbf{105}, 136805\relax
\mciteBstWouldAddEndPuncttrue
\mciteSetBstMidEndSepPunct{\mcitedefaultmidpunct}
{\mcitedefaultendpunct}{\mcitedefaultseppunct}\relax
\EndOfBibitem
\bibitem[Splendiani \emph{et~al.}(2010)Splendiani, Sun, Zhang, Li, Kim, Chim,
  Galli, and Wang]{Splendiani2010}
A.~Splendiani, L.~Sun, Y.~Zhang, T.~Li, J.~Kim, C.-Y. Chim, G.~Galli and
  F.~Wang, \emph{Nano letters}, 2010, \textbf{10}, 1271--1275\relax
\mciteBstWouldAddEndPuncttrue
\mciteSetBstMidEndSepPunct{\mcitedefaultmidpunct}
{\mcitedefaultendpunct}{\mcitedefaultseppunct}\relax
\EndOfBibitem
\bibitem[Chernikov \emph{et~al.}(2014)Chernikov, Berkelbach, Hill, Rigosi, Li,
  Aslan, Reichman, Hybertsen, and Heinz]{Chernikov2014}
A.~Chernikov, T.~C. Berkelbach, H.~M. Hill, A.~Rigosi, Y.~Li, O.~B. Aslan,
  D.~R. Reichman, M.~S. Hybertsen and T.~F. Heinz, \emph{Physical review
  letters}, 2014, \textbf{113}, 076802\relax
\mciteBstWouldAddEndPuncttrue
\mciteSetBstMidEndSepPunct{\mcitedefaultmidpunct}
{\mcitedefaultendpunct}{\mcitedefaultseppunct}\relax
\EndOfBibitem
\bibitem[Wang \emph{et~al.}(2018)Wang, Chernikov, Glazov, Heinz, Marie, Amand,
  and Urbaszek]{Wang2018}
G.~Wang, A.~Chernikov, M.~M. Glazov, T.~F. Heinz, X.~Marie, T.~Amand and
  B.~Urbaszek, \emph{Reviews of Modern Physics}, 2018, \textbf{90},
  021001\relax
\mciteBstWouldAddEndPuncttrue
\mciteSetBstMidEndSepPunct{\mcitedefaultmidpunct}
{\mcitedefaultendpunct}{\mcitedefaultseppunct}\relax
\EndOfBibitem
\bibitem[Cordovilla~Leon \emph{et~al.}(2018)Cordovilla~Leon, Li, Jang, Cheng,
  and Deotare]{Cordovilla2018}
D.~F. Cordovilla~Leon, Z.~Li, S.~W. Jang, C.-H. Cheng and P.~B. Deotare,
  \emph{Applied Physics Letters}, 2018, \textbf{113}, 252101\relax
\mciteBstWouldAddEndPuncttrue
\mciteSetBstMidEndSepPunct{\mcitedefaultmidpunct}
{\mcitedefaultendpunct}{\mcitedefaultseppunct}\relax
\EndOfBibitem
\bibitem[Ishihara \emph{et~al.}(1989)Ishihara, Takahashi, and
  Goto]{Ishihara1989}
T.~Ishihara, J.~Takahashi and T.~Goto, \emph{Solid state communications}, 1989,
  \textbf{69}, 933--936\relax
\mciteBstWouldAddEndPuncttrue
\mciteSetBstMidEndSepPunct{\mcitedefaultmidpunct}
{\mcitedefaultendpunct}{\mcitedefaultseppunct}\relax
\EndOfBibitem
\bibitem[Mitzi \emph{et~al.}(1994)Mitzi, Feild, Harrison, and Guloy]{Mitzi1994}
D.~B. Mitzi, C.~Feild, W.~Harrison and A.~Guloy, \emph{Nature}, 1994,
  \textbf{369}, 467--469\relax
\mciteBstWouldAddEndPuncttrue
\mciteSetBstMidEndSepPunct{\mcitedefaultmidpunct}
{\mcitedefaultendpunct}{\mcitedefaultseppunct}\relax
\EndOfBibitem
\bibitem[Becker \emph{et~al.}(2018)Becker, Vaxenburg, Nedelcu, Sercel, Shabaev,
  Mehl, Michopoulos, Lambrakos, Bernstein, Lyons,\emph{et~al.}]{Becker2018}
M.~A. Becker, R.~Vaxenburg, G.~Nedelcu, P.~C. Sercel, A.~Shabaev, M.~J. Mehl,
  J.~G. Michopoulos, S.~G. Lambrakos, N.~Bernstein, J.~L. Lyons \emph{et~al.},
  \emph{Nature}, 2018, \textbf{553}, 189--193\relax
\mciteBstWouldAddEndPuncttrue
\mciteSetBstMidEndSepPunct{\mcitedefaultmidpunct}
{\mcitedefaultendpunct}{\mcitedefaultseppunct}\relax
\EndOfBibitem
\bibitem[Belykh \emph{et~al.}(2019)Belykh, Yakovlev, Glazov, Grigoryev,
  Hussain, Rautert, Dirin, Kovalenko, and Bayer]{Belykh2019}
V.~V. Belykh, D.~R. Yakovlev, M.~M. Glazov, P.~S. Grigoryev, M.~Hussain,
  J.~Rautert, D.~N. Dirin, M.~V. Kovalenko and M.~Bayer, \emph{Nature
  communications}, 2019, \textbf{10}, 1--6\relax
\mciteBstWouldAddEndPuncttrue
\mciteSetBstMidEndSepPunct{\mcitedefaultmidpunct}
{\mcitedefaultendpunct}{\mcitedefaultseppunct}\relax
\EndOfBibitem
\bibitem[Ziegler \emph{et~al.}(2020)Ziegler, Zipfel, Meisinger, Menahem, Zhu,
  Taniguchi, Watanabe, Yaffe, Egger, and Chernikov]{Ziegler2020}
J.~D. Ziegler, J.~Zipfel, B.~Meisinger, M.~Menahem, X.~Zhu, T.~Taniguchi,
  K.~Watanabe, O.~Yaffe, D.~A. Egger and A.~Chernikov, \emph{Nano Letters},
  2020, \textbf{20}, 6674--6681\relax
\mciteBstWouldAddEndPuncttrue
\mciteSetBstMidEndSepPunct{\mcitedefaultmidpunct}
{\mcitedefaultendpunct}{\mcitedefaultseppunct}\relax
\EndOfBibitem
\bibitem[Magdaleno \emph{et~al.}(2021)Magdaleno, Seitz, Frising, de~la Cruz,
  Fern{\'a}ndez-Dom{\'\i}nguez, and Prins]{Magdaleno2021}
A.~J. Magdaleno, M.~Seitz, M.~Frising, A.~H. de~la Cruz, A.~I.
  Fern{\'a}ndez-Dom{\'\i}nguez and F.~Prins, \emph{Materials Horizons}, 2021,
  \textbf{8}, 639--644\relax
\mciteBstWouldAddEndPuncttrue
\mciteSetBstMidEndSepPunct{\mcitedefaultmidpunct}
{\mcitedefaultendpunct}{\mcitedefaultseppunct}\relax
\EndOfBibitem
\bibitem[Seitz \emph{et~al.}(2020)Seitz, Magdaleno, Alc{\'a}zar-Cano,
  Mel{\'e}ndez, Lubbers, Walraven, Pakdel, Prada, Delgado-Buscalioni, and
  Prins]{Seitz2020}
M.~Seitz, A.~J. Magdaleno, N.~Alc{\'a}zar-Cano, M.~Mel{\'e}ndez, T.~J. Lubbers,
  S.~W. Walraven, S.~Pakdel, E.~Prada, R.~Delgado-Buscalioni and F.~Prins,
  \emph{Nature communications}, 2020, \textbf{11}, 1--8\relax
\mciteBstWouldAddEndPuncttrue
\mciteSetBstMidEndSepPunct{\mcitedefaultmidpunct}
{\mcitedefaultendpunct}{\mcitedefaultseppunct}\relax
\EndOfBibitem
\bibitem[Konstantatos \emph{et~al.}(2012)Konstantatos, Badioli, Gaudreau,
  Osmond, Bernechea, De~Arquer, Gatti, and Koppens]{Konstantatos2012}
G.~Konstantatos, M.~Badioli, L.~Gaudreau, J.~Osmond, M.~Bernechea, F.~P.~G.
  De~Arquer, F.~Gatti and F.~H. Koppens, \emph{Nature nanotechnology}, 2012,
  \textbf{7}, 363--368\relax
\mciteBstWouldAddEndPuncttrue
\mciteSetBstMidEndSepPunct{\mcitedefaultmidpunct}
{\mcitedefaultendpunct}{\mcitedefaultseppunct}\relax
\EndOfBibitem
\bibitem[Baugher \emph{et~al.}(2014)Baugher, Churchill, Yang, and
  Jarillo-Herrero]{Baugher2014}
B.~W. Baugher, H.~O. Churchill, Y.~Yang and P.~Jarillo-Herrero, \emph{Nature
  nanotechnology}, 2014, \textbf{9}, 262--267\relax
\mciteBstWouldAddEndPuncttrue
\mciteSetBstMidEndSepPunct{\mcitedefaultmidpunct}
{\mcitedefaultendpunct}{\mcitedefaultseppunct}\relax
\EndOfBibitem
\bibitem[Xing \emph{et~al.}(2018)Xing, Zhao, Askerka, Quan, Gong, Zhao, Zhao,
  Tan, Long, Gao,\emph{et~al.}]{Xing2018}
J.~Xing, Y.~Zhao, M.~Askerka, L.~N. Quan, X.~Gong, W.~Zhao, J.~Zhao, H.~Tan,
  G.~Long, L.~Gao \emph{et~al.}, \emph{Nature communications}, 2018,
  \textbf{9}, 1--8\relax
\mciteBstWouldAddEndPuncttrue
\mciteSetBstMidEndSepPunct{\mcitedefaultmidpunct}
{\mcitedefaultendpunct}{\mcitedefaultseppunct}\relax
\EndOfBibitem
\bibitem[Yang \emph{et~al.}(2017)Yang, Fu, Zhang, Chen, and Li]{Yang2017}
S.~Yang, W.~Fu, Z.~Zhang, H.~Chen and C.-Z. Li, \emph{Journal of Materials
  Chemistry A}, 2017, \textbf{5}, 11462--11482\relax
\mciteBstWouldAddEndPuncttrue
\mciteSetBstMidEndSepPunct{\mcitedefaultmidpunct}
{\mcitedefaultendpunct}{\mcitedefaultseppunct}\relax
\EndOfBibitem
\bibitem[Smith \emph{et~al.}(2014)Smith, Hoke, Solis-Ibarra, McGehee, and
  Karunadasa]{Smith2014}
I.~C. Smith, E.~T. Hoke, D.~Solis-Ibarra, M.~D. McGehee and H.~I. Karunadasa,
  \emph{Angewandte Chemie International Edition}, 2014, \textbf{53},
  11232--11235\relax
\mciteBstWouldAddEndPuncttrue
\mciteSetBstMidEndSepPunct{\mcitedefaultmidpunct}
{\mcitedefaultendpunct}{\mcitedefaultseppunct}\relax
\EndOfBibitem
\bibitem[Yuan \emph{et~al.}(2017)Yuan, Wang, Zhu, Zhou, and Huang]{Yuan2017}
L.~Yuan, T.~Wang, T.~Zhu, M.~Zhou and L.~Huang, \emph{The journal of physical
  chemistry letters}, 2017, \textbf{8}, 3371--3379\relax
\mciteBstWouldAddEndPuncttrue
\mciteSetBstMidEndSepPunct{\mcitedefaultmidpunct}
{\mcitedefaultendpunct}{\mcitedefaultseppunct}\relax
\EndOfBibitem
\bibitem[Cadiz \emph{et~al.}(2018)Cadiz, Robert, Courtade, Manca, Martinelli,
  Taniguchi, Watanabe, Amand, Rowe, Paget,\emph{et~al.}]{Cadiz2018}
F.~Cadiz, C.~Robert, E.~Courtade, M.~Manca, L.~Martinelli, T.~Taniguchi,
  K.~Watanabe, T.~Amand, A.~Rowe, D.~Paget \emph{et~al.}, \emph{Applied Physics
  Letters}, 2018, \textbf{112}, 152106\relax
\mciteBstWouldAddEndPuncttrue
\mciteSetBstMidEndSepPunct{\mcitedefaultmidpunct}
{\mcitedefaultendpunct}{\mcitedefaultseppunct}\relax
\EndOfBibitem
\bibitem[Kulig \emph{et~al.}(2018)Kulig, Zipfel, Nagler, Blanter, Sch{\"u}ller,
  Korn, Paradiso, Glazov, and Chernikov]{Kulig2018}
M.~Kulig, J.~Zipfel, P.~Nagler, S.~Blanter, C.~Sch{\"u}ller, T.~Korn,
  N.~Paradiso, M.~M. Glazov and A.~Chernikov, \emph{Physical review letters},
  2018, \textbf{120}, 207401\relax
\mciteBstWouldAddEndPuncttrue
\mciteSetBstMidEndSepPunct{\mcitedefaultmidpunct}
{\mcitedefaultendpunct}{\mcitedefaultseppunct}\relax
\EndOfBibitem
\bibitem[Bardeen(2014)]{Bardeen2014}
C.~J. Bardeen, \emph{Annual review of physical chemistry}, 2014, \textbf{65},
  127--148\relax
\mciteBstWouldAddEndPuncttrue
\mciteSetBstMidEndSepPunct{\mcitedefaultmidpunct}
{\mcitedefaultendpunct}{\mcitedefaultseppunct}\relax
\EndOfBibitem
\bibitem[Dong \emph{et~al.}(2015)Dong, Fang, Shao, Mulligan, Qiu, Cao, and
  Huang]{Dong2015}
Q.~Dong, Y.~Fang, Y.~Shao, P.~Mulligan, J.~Qiu, L.~Cao and J.~Huang,
  \emph{Science}, 2015, \textbf{347}, 967--970\relax
\mciteBstWouldAddEndPuncttrue
\mciteSetBstMidEndSepPunct{\mcitedefaultmidpunct}
{\mcitedefaultendpunct}{\mcitedefaultseppunct}\relax
\EndOfBibitem
\bibitem[Kurilovich \emph{et~al.}(2020)Kurilovich, Mantsevich, Stevenson,
  Chechkin, and Palyulin]{Kurilovich2020}
A.~A. Kurilovich, V.~N. Mantsevich, K.~J. Stevenson, A.~V. Chechkin and V.~V.
  Palyulin, \emph{Physical Chemistry Chemical Physics}, 2020, \textbf{22},
  24686--24696\relax
\mciteBstWouldAddEndPuncttrue
\mciteSetBstMidEndSepPunct{\mcitedefaultmidpunct}
{\mcitedefaultendpunct}{\mcitedefaultseppunct}\relax
\EndOfBibitem
\bibitem[Kurilovich \emph{et~al.}(2021)Kurilovich, Mantsevich, Stevenson,
  Chechkin, and Palyulin]{Kurilovich2021}
A.~Kurilovich, V.~Mantsevich, K.~Stevenson, A.~Chechkin and V.~Palyulin,
  Journal of Physics: Conference Series, 2021, p. 012103\relax
\mciteBstWouldAddEndPuncttrue
\mciteSetBstMidEndSepPunct{\mcitedefaultmidpunct}
{\mcitedefaultendpunct}{\mcitedefaultseppunct}\relax
\EndOfBibitem
\bibitem[Glazov(2019)]{Glazov2019}
M.~Glazov, \emph{Physical Review B}, 2019, \textbf{100}, 045426\relax
\mciteBstWouldAddEndPuncttrue
\mciteSetBstMidEndSepPunct{\mcitedefaultmidpunct}
{\mcitedefaultendpunct}{\mcitedefaultseppunct}\relax
\EndOfBibitem
\bibitem[Shaw \emph{et~al.}(2008)Shaw, Ruseckas, and Samuel]{Shaw2008}
P.~E. Shaw, A.~Ruseckas and I.~D. Samuel, \emph{Advanced Materials}, 2008,
  \textbf{20}, 3516--3520\relax
\mciteBstWouldAddEndPuncttrue
\mciteSetBstMidEndSepPunct{\mcitedefaultmidpunct}
{\mcitedefaultendpunct}{\mcitedefaultseppunct}\relax
\EndOfBibitem
\bibitem[Stevens \emph{et~al.}(2001)Stevens, Silva, Russell, and
  Friend]{Stevens2001}
M.~A. Stevens, C.~Silva, D.~M. Russell and R.~H. Friend, \emph{Physical Review
  B}, 2001, \textbf{63}, 165213\relax
\mciteBstWouldAddEndPuncttrue
\mciteSetBstMidEndSepPunct{\mcitedefaultmidpunct}
{\mcitedefaultendpunct}{\mcitedefaultseppunct}\relax
\EndOfBibitem
\bibitem[Del Pozo-Zamudio \emph{et~al.}(2015)Del Pozo-Zamudio, Schwarz, Sich,
  Akimov, Bayer, Schofield, Chekhovich, Robinson, Kay,
  Kolosov,\emph{et~al.}]{Zamudio2015}
O.~Del Pozo-Zamudio, S.~Schwarz, M.~Sich, I.~Akimov, M.~Bayer, R.~Schofield,
  E.~Chekhovich, B.~Robinson, N.~Kay, O.~Kolosov \emph{et~al.}, \emph{2D
  Materials}, 2015, \textbf{2}, 035010\relax
\mciteBstWouldAddEndPuncttrue
\mciteSetBstMidEndSepPunct{\mcitedefaultmidpunct}
{\mcitedefaultendpunct}{\mcitedefaultseppunct}\relax
\EndOfBibitem
\bibitem[Akselrod \emph{et~al.}(2014)Akselrod, Deotare, Thompson, Lee, Tisdale,
  Baldo, Menon, and Bulovi{\'c}]{Akselrod2014}
G.~M. Akselrod, P.~B. Deotare, N.~J. Thompson, J.~Lee, W.~A. Tisdale, M.~A.
  Baldo, V.~M. Menon and V.~Bulovi{\'c}, \emph{Nature communications}, 2014,
  \textbf{5}, 1--8\relax
\mciteBstWouldAddEndPuncttrue
\mciteSetBstMidEndSepPunct{\mcitedefaultmidpunct}
{\mcitedefaultendpunct}{\mcitedefaultseppunct}\relax
\EndOfBibitem
\bibitem[Han \emph{et~al.}(2018)Han, Robert, Courtade, Manca, Shree, Amand,
  Renucci, Taniguchi, Watanabe, Marie,\emph{et~al.}]{Han2018}
B.~Han, C.~Robert, E.~Courtade, M.~Manca, S.~Shree, T.~Amand, P.~Renucci,
  T.~Taniguchi, K.~Watanabe, X.~Marie \emph{et~al.}, \emph{Physical Review X},
  2018, \textbf{8}, 031073\relax
\mciteBstWouldAddEndPuncttrue
\mciteSetBstMidEndSepPunct{\mcitedefaultmidpunct}
{\mcitedefaultendpunct}{\mcitedefaultseppunct}\relax
\EndOfBibitem
\bibitem[Ginsberg and Tisdale(2020)]{Ginsberg2020}
N.~S. Ginsberg and W.~A. Tisdale, \emph{Annual review of physical chemistry},
  2020, \textbf{71}, 1--30\relax
\mciteBstWouldAddEndPuncttrue
\mciteSetBstMidEndSepPunct{\mcitedefaultmidpunct}
{\mcitedefaultendpunct}{\mcitedefaultseppunct}\relax
\EndOfBibitem
\bibitem[Markov \emph{et~al.}(2005)Markov, Tanase, Blom, and
  Wildeman]{Markov2005}
D.~Markov, C.~Tanase, P.~Blom and J.~Wildeman, \emph{Physical Review B}, 2005,
  \textbf{72}, 045217\relax
\mciteBstWouldAddEndPuncttrue
\mciteSetBstMidEndSepPunct{\mcitedefaultmidpunct}
{\mcitedefaultendpunct}{\mcitedefaultseppunct}\relax
\EndOfBibitem
\bibitem[Mikhnenko \emph{et~al.}(2008)Mikhnenko, Cordella, Sieval, Hummelen,
  Blom, and Loi]{Mikhnenko2008}
O.~Mikhnenko, F.~Cordella, A.~Sieval, J.~Hummelen, P.~Blom and M.~Loi,
  \emph{The Journal of Physical Chemistry B}, 2008, \textbf{112},
  11601--11604\relax
\mciteBstWouldAddEndPuncttrue
\mciteSetBstMidEndSepPunct{\mcitedefaultmidpunct}
{\mcitedefaultendpunct}{\mcitedefaultseppunct}\relax
\EndOfBibitem
\bibitem[Kholmicheva \emph{et~al.}(2015)Kholmicheva, Moroz, Bastola,
  Razgoniaeva, Bocanegra, Shaughnessy, Porach, Khon, and
  Zamkov]{Kholmicheva2015}
N.~Kholmicheva, P.~Moroz, E.~Bastola, N.~Razgoniaeva, J.~Bocanegra,
  M.~Shaughnessy, Z.~Porach, D.~Khon and M.~Zamkov, \emph{ACS nano}, 2015,
  \textbf{9}, 2926--2937\relax
\mciteBstWouldAddEndPuncttrue
\mciteSetBstMidEndSepPunct{\mcitedefaultmidpunct}
{\mcitedefaultendpunct}{\mcitedefaultseppunct}\relax
\EndOfBibitem
\bibitem[Kenkre(1983)]{Kenkre1983}
V.~Kenkre, \emph{Journal of Statistical Physics}, 1983, \textbf{30},
  293--303\relax
\mciteBstWouldAddEndPuncttrue
\mciteSetBstMidEndSepPunct{\mcitedefaultmidpunct}
{\mcitedefaultendpunct}{\mcitedefaultseppunct}\relax
\EndOfBibitem
\bibitem[Heijs \emph{et~al.}(2005)Heijs, Malyshev, and Knoester]{Heijs2005}
D.~Heijs, V.~Malyshev and J.~Knoester, \emph{Physical review letters}, 2005,
  \textbf{95}, 177402\relax
\mciteBstWouldAddEndPuncttrue
\mciteSetBstMidEndSepPunct{\mcitedefaultmidpunct}
{\mcitedefaultendpunct}{\mcitedefaultseppunct}\relax
\EndOfBibitem
\bibitem[Akselrod \emph{et~al.}(2014)Akselrod, Prins, Poulikakos, Lee, Weidman,
  Mork, Willard, Bulovic, and Tisdale]{Akselrod2014_1}
G.~M. Akselrod, F.~Prins, L.~V. Poulikakos, E.~M. Lee, M.~C. Weidman, A.~J.
  Mork, A.~P. Willard, V.~Bulovic and W.~A. Tisdale, \emph{Nano letters}, 2014,
  \textbf{14}, 3556--3562\relax
\mciteBstWouldAddEndPuncttrue
\mciteSetBstMidEndSepPunct{\mcitedefaultmidpunct}
{\mcitedefaultendpunct}{\mcitedefaultseppunct}\relax
\EndOfBibitem
\bibitem[Miyazaki and Kinoshita(2012)]{Miyazaki2012}
J.~Miyazaki and S.~Kinoshita, \emph{Physical Review B}, 2012, \textbf{86},
  035303\relax
\mciteBstWouldAddEndPuncttrue
\mciteSetBstMidEndSepPunct{\mcitedefaultmidpunct}
{\mcitedefaultendpunct}{\mcitedefaultseppunct}\relax
\EndOfBibitem
\bibitem[Vlaming \emph{et~al.}(2013)Vlaming, Malyshev, Eisfeld, and
  Knoester]{vlaming2013}
S.~Vlaming, V.~Malyshev, A.~Eisfeld and J.~Knoester, \emph{The Journal of
  chemical physics}, 2013, \textbf{138}, 214316\relax
\mciteBstWouldAddEndPuncttrue
\mciteSetBstMidEndSepPunct{\mcitedefaultmidpunct}
{\mcitedefaultendpunct}{\mcitedefaultseppunct}\relax
\EndOfBibitem
\bibitem[Kurilovich \emph{et~al.}(2022)Kurilovich, Mantsevich, Mardoukhi,
  Stevenson, Chechkin, and Palyulin]{Kurilovich2022}
A.~A. Kurilovich, V.~N. Mantsevich, Y.~Mardoukhi, K.~J. Stevenson, A.~V.
  Chechkin and V.~V. Palyulin, \emph{Physical Chemistry Chemical Physics},
  2022, \textbf{24}, 13941--13950\relax
\mciteBstWouldAddEndPuncttrue
\mciteSetBstMidEndSepPunct{\mcitedefaultmidpunct}
{\mcitedefaultendpunct}{\mcitedefaultseppunct}\relax
\EndOfBibitem
\bibitem[Lunt \emph{et~al.}(2009)Lunt, Giebink, Belak, Benziger, and
  Forrest]{Lunt2009}
R.~R. Lunt, N.~C. Giebink, A.~A. Belak, J.~B. Benziger and S.~R. Forrest,
  \emph{Journal of Applied Physics}, 2009, \textbf{105}, 053711\relax
\mciteBstWouldAddEndPuncttrue
\mciteSetBstMidEndSepPunct{\mcitedefaultmidpunct}
{\mcitedefaultendpunct}{\mcitedefaultseppunct}\relax
\EndOfBibitem
\bibitem[Lee and Tisdale(2015)]{Lee2015}
E.~M. Lee and W.~A. Tisdale, \emph{The Journal of Physical Chemistry C}, 2015,
  \textbf{119}, 9005--9015\relax
\mciteBstWouldAddEndPuncttrue
\mciteSetBstMidEndSepPunct{\mcitedefaultmidpunct}
{\mcitedefaultendpunct}{\mcitedefaultseppunct}\relax
\EndOfBibitem
\bibitem[Lin \emph{et~al.}(2016)Lin, Carvalho, Kahn, Lv, Rao, Terrones,
  Pimenta, and Terrones]{Lin2016}
Z.~Lin, B.~R. Carvalho, E.~Kahn, R.~Lv, R.~Rao, H.~Terrones, M.~A. Pimenta and
  M.~Terrones, \emph{2D Materials}, 2016, \textbf{3}, 022002\relax
\mciteBstWouldAddEndPuncttrue
\mciteSetBstMidEndSepPunct{\mcitedefaultmidpunct}
{\mcitedefaultendpunct}{\mcitedefaultseppunct}\relax
\EndOfBibitem
\bibitem[Kolobkova \emph{et~al.}(2021)Kolobkova, Kuznetsova, and
  Nikonorov]{Kolobkova2021}
E.~Kolobkova, M.~Kuznetsova and N.~Nikonorov, \emph{Journal of Non-Crystalline
  Solids}, 2021, \textbf{563}, 120811\relax
\mciteBstWouldAddEndPuncttrue
\mciteSetBstMidEndSepPunct{\mcitedefaultmidpunct}
{\mcitedefaultendpunct}{\mcitedefaultseppunct}\relax
\EndOfBibitem
\bibitem[Doerries \emph{et~al.}(2022)Doerries, Chechkin, Schumer, and
  Metzler]{alekseiralf2022}
T.~J. Doerries, A.~V. Chechkin, R.~Schumer and R.~Metzler, \emph{Physical
  Review E}, 2022, \textbf{105}, 014105\relax
\mciteBstWouldAddEndPuncttrue
\mciteSetBstMidEndSepPunct{\mcitedefaultmidpunct}
{\mcitedefaultendpunct}{\mcitedefaultseppunct}\relax
\EndOfBibitem
\bibitem[Coats and Smith(1964)]{coats1964}
K.~Coats and B.~Smith, \emph{Society of petroleum engineers journal}, 1964,
  \textbf{4}, 73--84\relax
\mciteBstWouldAddEndPuncttrue
\mciteSetBstMidEndSepPunct{\mcitedefaultmidpunct}
{\mcitedefaultendpunct}{\mcitedefaultseppunct}\relax
\EndOfBibitem
\bibitem[Schumer \emph{et~al.}(2003)Schumer, Benson, Meerschaert, and
  Baeumer]{schumer2003}
R.~Schumer, D.~A. Benson, M.~M. Meerschaert and B.~Baeumer, \emph{Water
  Resources Research}, 2003, \textbf{39}, \relax
\mciteBstWouldAddEndPuncttrue
\mciteSetBstMidEndSepPunct{\mcitedefaultmidpunct}
{\mcitedefaultendpunct}{\mcitedefaultseppunct}\relax
\EndOfBibitem
\bibitem[Igaev \emph{et~al.}(2014)Igaev, Janning, S{\"u}ndermann, Niewidok,
  Brandt, and Junge]{igaev2014}
M.~Igaev, D.~Janning, F.~S{\"u}ndermann, B.~Niewidok, R.~Brandt and W.~Junge,
  \emph{Biophysical journal}, 2014, \textbf{107}, 2567--2578\relax
\mciteBstWouldAddEndPuncttrue
\mciteSetBstMidEndSepPunct{\mcitedefaultmidpunct}
{\mcitedefaultendpunct}{\mcitedefaultseppunct}\relax
\EndOfBibitem
\bibitem[Doerries \emph{et~al.}(2022)Doerries, Chechkin, and
  Metzler]{doerries2022apparent}
T.~J. Doerries, A.~V. Chechkin and R.~Metzler, \emph{Journal of the Royal
  Society Interface}, 2022, \textbf{19}, 20220233\relax
\mciteBstWouldAddEndPuncttrue
\mciteSetBstMidEndSepPunct{\mcitedefaultmidpunct}
{\mcitedefaultendpunct}{\mcitedefaultseppunct}\relax
\EndOfBibitem
\bibitem[Mora and Pomeau(2018)]{mora2018}
S.~Mora and Y.~Pomeau, \emph{Physical Review E}, 2018, \textbf{98},
  040101\relax
\mciteBstWouldAddEndPuncttrue
\mciteSetBstMidEndSepPunct{\mcitedefaultmidpunct}
{\mcitedefaultendpunct}{\mcitedefaultseppunct}\relax
\EndOfBibitem
\bibitem[Doerries \emph{et~al.}(2023)Doerries, Metzler, and
  Chechkin]{doerries2023emergent}
T.~Doerries, R.~Metzler and A.~Chechkin, \emph{New Journal of Physics}, 2023,
  \textbf{25}, 063009\relax
\mciteBstWouldAddEndPuncttrue
\mciteSetBstMidEndSepPunct{\mcitedefaultmidpunct}
{\mcitedefaultendpunct}{\mcitedefaultseppunct}\relax
\EndOfBibitem
\bibitem[Rosati \emph{et~al.}(2020)Rosati, Perea-Caus{\'{\i}}n, Brem, and
  Malic]{Rosati2020}
R.~Rosati, R.~Perea-Caus{\'{\i}}n, S.~Brem and E.~Malic, \emph{Nanoscale},
  2020, \textbf{12}, 356--363\relax
\mciteBstWouldAddEndPuncttrue
\mciteSetBstMidEndSepPunct{\mcitedefaultmidpunct}
{\mcitedefaultendpunct}{\mcitedefaultseppunct}\relax
\EndOfBibitem
\bibitem[Rosati \emph{et~al.}(2021)Rosati, Wagner, Brem, Perea-Caus{\'\i}n,
  Ziegler, Zipfel, Taniguchi, Watanabe, Chernikov, and Malic]{Rosati2021}
R.~Rosati, K.~Wagner, S.~Brem, R.~Perea-Caus{\'\i}n, J.~D. Ziegler, J.~Zipfel,
  T.~Taniguchi, K.~Watanabe, A.~Chernikov and E.~Malic, \emph{Nanoscale}, 2021,
  \textbf{13}, 19966--19972\relax
\mciteBstWouldAddEndPuncttrue
\mciteSetBstMidEndSepPunct{\mcitedefaultmidpunct}
{\mcitedefaultendpunct}{\mcitedefaultseppunct}\relax
\EndOfBibitem
\bibitem[Zhao \emph{et~al.}(2005)Zhao, Dal~Don, Schwartz, and Kalt]{Zhao2005}
H.~Zhao, B.~Dal~Don, G.~Schwartz and H.~Kalt, \emph{Phys. Rev. Lett.}, 2005,
  \textbf{94}, 137402\relax
\mciteBstWouldAddEndPuncttrue
\mciteSetBstMidEndSepPunct{\mcitedefaultmidpunct}
{\mcitedefaultendpunct}{\mcitedefaultseppunct}\relax
\EndOfBibitem
\bibitem[Nakade \emph{et~al.}(2021)Nakade, Kanki, Tanaka, and
  Petrosky]{nakade2021anomalous}
S.~Nakade, K.~Kanki, S.~Tanaka and T.~Petrosky, \emph{Symmetry}, 2021,
  \textbf{13}, 506\relax
\mciteBstWouldAddEndPuncttrue
\mciteSetBstMidEndSepPunct{\mcitedefaultmidpunct}
{\mcitedefaultendpunct}{\mcitedefaultseppunct}\relax
\EndOfBibitem
\bibitem[Beret \emph{et~al.}(2023)Beret, Ren, Robert, Foussat, Renucci,
  Lagarde, Balocchi, Amand, Urbaszek,
  Watanabe,\emph{et~al.}]{beret2023nonlinear}
D.~Beret, L.~Ren, C.~Robert, L.~Foussat, P.~Renucci, D.~Lagarde, A.~Balocchi,
  T.~Amand, B.~Urbaszek, K.~Watanabe \emph{et~al.}, \emph{Physical Review B},
  2023, \textbf{107}, 045420\relax
\mciteBstWouldAddEndPuncttrue
\mciteSetBstMidEndSepPunct{\mcitedefaultmidpunct}
{\mcitedefaultendpunct}{\mcitedefaultseppunct}\relax
\EndOfBibitem
\bibitem[Philbin and Rabani(2018)]{Philbin2018}
J.~P. Philbin and E.~Rabani, \emph{Nano Letters}, 2018, \textbf{18},
  7889--7895\relax
\mciteBstWouldAddEndPuncttrue
\mciteSetBstMidEndSepPunct{\mcitedefaultmidpunct}
{\mcitedefaultendpunct}{\mcitedefaultseppunct}\relax
\EndOfBibitem
\bibitem[Lewis \emph{et~al.}(2006)Lewis, Ruseckas, Gaudin, Webster, Burn, and
  Samuel]{Lewis2006}
A.~Lewis, A.~Ruseckas, O.~Gaudin, G.~Webster, P.~Burn and I.~Samuel,
  \emph{Organic Electronics}, 2006, \textbf{7}, 452--456\relax
\mciteBstWouldAddEndPuncttrue
\mciteSetBstMidEndSepPunct{\mcitedefaultmidpunct}
{\mcitedefaultendpunct}{\mcitedefaultseppunct}\relax
\EndOfBibitem
\bibitem[Maniloff \emph{et~al.}(1997)Maniloff, Klimov, and
  McBranch]{Maniloff1997}
E.~S. Maniloff, V.~I. Klimov and D.~W. McBranch, \emph{Phys. Rev. B}, 1997,
  \textbf{56}, 1876--1881\relax
\mciteBstWouldAddEndPuncttrue
\mciteSetBstMidEndSepPunct{\mcitedefaultmidpunct}
{\mcitedefaultendpunct}{\mcitedefaultseppunct}\relax
\EndOfBibitem
\bibitem[Dogariu \emph{et~al.}(1998)Dogariu, Vacar, and Heeger]{Dogariu1998}
A.~Dogariu, D.~Vacar and A.~J. Heeger, \emph{Phys. Rev. B}, 1998, \textbf{58},
  10218--10224\relax
\mciteBstWouldAddEndPuncttrue
\mciteSetBstMidEndSepPunct{\mcitedefaultmidpunct}
{\mcitedefaultendpunct}{\mcitedefaultseppunct}\relax
\EndOfBibitem
\bibitem[Baghani \emph{et~al.}(2015)Baghani, O'Leary, Fedin, Talapin, and
  Pelton]{Baghani2015}
E.~Baghani, S.~K. O'Leary, I.~Fedin, D.~V. Talapin and M.~Pelton, \emph{The
  Journal of Physical Chemistry Letters}, 2015, \textbf{6}, 1032--1036\relax
\mciteBstWouldAddEndPuncttrue
\mciteSetBstMidEndSepPunct{\mcitedefaultmidpunct}
{\mcitedefaultendpunct}{\mcitedefaultseppunct}\relax
\EndOfBibitem
\bibitem[Smirnov \emph{et~al.}(2021)Smirnov, Golinskaya, Saidzhonov, Vasiliev,
  Mantsevich, and Dneprovskii]{Smirnov2021}
A.~M. Smirnov, A.~D. Golinskaya, B.~M. Saidzhonov, R.~B. Vasiliev, V.~N.
  Mantsevich and V.~S. Dneprovskii, \emph{Journal of Luminescence}, 2021,
  \textbf{229}, 117682\relax
\mciteBstWouldAddEndPuncttrue
\mciteSetBstMidEndSepPunct{\mcitedefaultmidpunct}
{\mcitedefaultendpunct}{\mcitedefaultseppunct}\relax
\EndOfBibitem
\bibitem[Kajino \emph{et~al.}(2021)Kajino, Sakanashi, Aoki, Watanabe,
  Taniguchi, Oto, and Yamada]{Kajino2021}
Y.~Kajino, K.~Sakanashi, N.~Aoki, K.~Watanabe, T.~Taniguchi, K.~Oto and
  Y.~Yamada, \emph{Phys. Rev. B}, 2021, \textbf{103}, L241410\relax
\mciteBstWouldAddEndPuncttrue
\mciteSetBstMidEndSepPunct{\mcitedefaultmidpunct}
{\mcitedefaultendpunct}{\mcitedefaultseppunct}\relax
\EndOfBibitem
\bibitem[Hoshi \emph{et~al.}(2017)Hoshi, Kuroda, Okada, Moriya, Masubuchi,
  Watanabe, Taniguchi, Kitaura, and Machida]{Hoshi2017}
Y.~Hoshi, T.~Kuroda, M.~Okada, R.~Moriya, S.~Masubuchi, K.~Watanabe,
  T.~Taniguchi, R.~Kitaura and T.~Machida, \emph{Phys. Rev. B}, 2017,
  \textbf{95}, 241403\relax
\mciteBstWouldAddEndPuncttrue
\mciteSetBstMidEndSepPunct{\mcitedefaultmidpunct}
{\mcitedefaultendpunct}{\mcitedefaultseppunct}\relax
\EndOfBibitem
\bibitem[Deng \emph{et~al.}(2020)Deng, Shi, Yuan, Jin, Dou, and
  Huang]{Deng2020}
S.~Deng, E.~Shi, L.~Yuan, L.~Jin, L.~Dou and L.~Huang, \emph{Nature
  Communications}, 2020, \textbf{11}, 664\relax
\mciteBstWouldAddEndPuncttrue
\mciteSetBstMidEndSepPunct{\mcitedefaultmidpunct}
{\mcitedefaultendpunct}{\mcitedefaultseppunct}\relax
\EndOfBibitem
\bibitem[Hinterding \emph{et~al.}(2021)Hinterding, Salzmann, Vonk,
  Vanmaekelbergh, Weckhuysen, Hutter, and Rabouw]{hinterding2021single}
S.~O. Hinterding, B.~B. Salzmann, S.~J. Vonk, D.~Vanmaekelbergh, B.~M.
  Weckhuysen, E.~M. Hutter and F.~T. Rabouw, \emph{ACS nano}, 2021,
  \textbf{15}, 7216--7225\relax
\mciteBstWouldAddEndPuncttrue
\mciteSetBstMidEndSepPunct{\mcitedefaultmidpunct}
{\mcitedefaultendpunct}{\mcitedefaultseppunct}\relax
\EndOfBibitem
\bibitem[Metzler \emph{et~al.}(2014)Metzler, Jeon, Cherstvy, and
  Barkai]{AndreyPCCP2014}
R.~Metzler, J.-H. Jeon, A.~G. Cherstvy and E.~Barkai, \emph{Physical Chemistry
  Chemical Physics}, 2014, \textbf{16}, 24128--24164\relax
\mciteBstWouldAddEndPuncttrue
\mciteSetBstMidEndSepPunct{\mcitedefaultmidpunct}
{\mcitedefaultendpunct}{\mcitedefaultseppunct}\relax
\EndOfBibitem
\bibitem[Defaveri \emph{et~al.}(2023)Defaveri, dos Santos, Kessler, Barkai, and
  Anteneodo]{Eli2023}
L.~Defaveri, M.~dos Santos, D.~Kessler, E.~Barkai and C.~Anteneodo,
  \emph{arXiv:2304.08834v1}, 2023\relax
\mciteBstWouldAddEndPuncttrue
\mciteSetBstMidEndSepPunct{\mcitedefaultmidpunct}
{\mcitedefaultendpunct}{\mcitedefaultseppunct}\relax
\EndOfBibitem
\bibitem[Bellour \emph{et~al.}(2002)Bellour, Skouri, Munch, and
  H{\'e}braud]{bellour2002brownian}
M.~Bellour, M.~Skouri, J.-P. Munch and P.~H{\'e}braud, \emph{The European
  Physical Journal E}, 2002, \textbf{8}, 431--436\relax
\mciteBstWouldAddEndPuncttrue
\mciteSetBstMidEndSepPunct{\mcitedefaultmidpunct}
{\mcitedefaultendpunct}{\mcitedefaultseppunct}\relax
\EndOfBibitem
\bibitem[Galvan-Miyoshi \emph{et~al.}(2008)Galvan-Miyoshi, Delgado, and
  Castillo]{galvan2008diffusing}
J.~Galvan-Miyoshi, J.~Delgado and R.~Castillo, \emph{The European Physical
  Journal E}, 2008, \textbf{26}, 369--377\relax
\mciteBstWouldAddEndPuncttrue
\mciteSetBstMidEndSepPunct{\mcitedefaultmidpunct}
{\mcitedefaultendpunct}{\mcitedefaultseppunct}\relax
\EndOfBibitem
\bibitem[Jeon \emph{et~al.}(2013)Jeon, Leijnse, Oddershede, and
  Metzler]{jeon2013anomalous}
J.-H. Jeon, N.~Leijnse, L.~B. Oddershede and R.~Metzler, \emph{New Journal of
  Physics}, 2013, \textbf{15}, 045011\relax
\mciteBstWouldAddEndPuncttrue
\mciteSetBstMidEndSepPunct{\mcitedefaultmidpunct}
{\mcitedefaultendpunct}{\mcitedefaultseppunct}\relax
\EndOfBibitem
\bibitem[{\AA}berg and Poolman(2021)]{aaberg2021glass}
C.~{\AA}berg and B.~Poolman, \emph{Biophysical Journal}, 2021, \textbf{120},
  2355--2366\relax
\mciteBstWouldAddEndPuncttrue
\mciteSetBstMidEndSepPunct{\mcitedefaultmidpunct}
{\mcitedefaultendpunct}{\mcitedefaultseppunct}\relax
\EndOfBibitem
\bibitem[Corci \emph{et~al.}(2023)Corci, Hooiveld, Dolga, and
  {\AA}berg]{corci2023extending}
B.~Corci, O.~Hooiveld, A.~M. Dolga and C.~{\AA}berg, \emph{Soft Matter}, 2023,
  \textbf{19}, 2529--2538\relax
\mciteBstWouldAddEndPuncttrue
\mciteSetBstMidEndSepPunct{\mcitedefaultmidpunct}
{\mcitedefaultendpunct}{\mcitedefaultseppunct}\relax
\EndOfBibitem
\bibitem[Wietek \emph{et~al.}(2023)Wietek, Florian, G{\"o}ser, Taniguchi,
  Watanabe, H{\"o}gele, Glazov, Steinhoff, and Chernikov]{wietek2023non}
E.~Wietek, M.~Florian, J.~M. G{\"o}ser, T.~Taniguchi, K.~Watanabe,
  A.~H{\"o}gele, M.~M. Glazov, A.~Steinhoff and A.~Chernikov, \emph{arXiv
  preprint arXiv:2306.12339}, 2023\relax
\mciteBstWouldAddEndPuncttrue
\mciteSetBstMidEndSepPunct{\mcitedefaultmidpunct}
{\mcitedefaultendpunct}{\mcitedefaultseppunct}\relax
\EndOfBibitem
\bibitem[Gorenflo \emph{et~al.}(2020)Gorenflo, Kilbas, Mainardi, and
  Rogosin]{gorenflo2020}
R.~Gorenflo, A.~Kilbas, F.~Mainardi and S.~Rogosin, \emph{Mittag-Leffler
  Functions, Related Topics and Applications}, Springer Berlin Heidelberg,
  2020\relax
\mciteBstWouldAddEndPuncttrue
\mciteSetBstMidEndSepPunct{\mcitedefaultmidpunct}
{\mcitedefaultendpunct}{\mcitedefaultseppunct}\relax
\EndOfBibitem
\bibitem[Haubold \emph{et~al.}(2011)Haubold, Mathai, and Saxena]{Haubold2011}
H.~J. Haubold, A.~M. Mathai and R.~K. Saxena, \emph{Journal of applied
  mathematics}, 2011,  298628\relax
\mciteBstWouldAddEndPuncttrue
\mciteSetBstMidEndSepPunct{\mcitedefaultmidpunct}
{\mcitedefaultendpunct}{\mcitedefaultseppunct}\relax
\EndOfBibitem
\end{mcitethebibliography}
\end{document}